\documentclass[12pt]{article}
\usepackage[utf8]{inputenc}
\usepackage[english]{babel}
\usepackage{url}
\usepackage{algorithm} 
\usepackage{algpseudocode}
\usepackage{subcaption}
\newcommand{\argmax}{arg\,max}
\usepackage{mathtools}
\usepackage{enumitem}
\usepackage{rotating}
\usepackage{tabls}
\usepackage[utf8]{inputenc}
\usepackage{siunitx}
\usepackage{color}
\usepackage[utf8]{inputenc}
\usepackage{natbib}

    \newcounter{lastnote}

\begin{document}
\title{Generalized Integrative Principal Component Analysis for Multi-Type Data with Block-Wise Missing Structure}

\author{HUICHEN ZHU*, GEN LI*, ERIC F. LOCK$^\dagger$\\
\textit{* The Department of Biostatistics, Columbia University, New York, NY, USA and}\\
\textit{$^\dagger$ The Division of Biostatistics, University of Minnesota, Minneapolis, MN, USA}
\\
{hz2366@cumc.columbia.edu}}
\date{}
\markboth%
{Huichen Zhu, Gen Li, Eric F. Lock}
{Generalized Integrative PCA}
\baselineskip24pt
\maketitle

\begin{abstract}
{High-dimensional multi-source data are encountered in many fields.  
Despite recent developments on the integrative dimension reduction of such data, most existing methods cannot easily accommodate data of multiple types (e.g., binary or count-valued).
Moreover, multi-source data often have block-wise missing structure, i.e., data in one or more sources may be completely unobserved for a sample.
The heterogeneous data types and presence of block-wise missing data pose significant challenges to the integration of multi-source data and further statistical analyses.
In this paper, we develop a low-rank method, called {\em Generalized Integrative Principal Component Analysis} (GIPCA), for the simultaneous dimension reduction and imputation of multi-source block-wise missing data, where different sources may have different data types.
We also devise an adapted BIC criterion for rank estimation.
Comprehensive simulation studies demonstrate the efficacy of the proposed method in terms of rank estimation, signal recovery, and missing data imputation.
We apply GIPCA to a mortality study. 
We achieve accurate block-wise missing data imputation and identify intriguing latent mortality rate patterns with sociological relevance.
} 
\end{abstract}

\section{Introduction}
\label{sec1}

With technological developments, data acquisition becomes easier and cheaper. 
In numerous studies, people collect data from multiple sources on the same group of objects, obtaining the so-called \emph{multi-source} (or multi-view) data.
The analysis of multi-source data presents many challenges.  One major challenge is the coexistence of heterogeneous data types in different data sources, such as continuous, binary, and count-valued data.
For instance, in genomic studies, data at different molecular levels such as RNA sequencing and DNA methylation data are collected from the same samples.
The next-generation RNA sequencing data typically take count values, while DNA methylation data are usually in the form of proportions between $0$ and $1$.
In addition to the diversity of data types, another challenge is the presence of \emph{block-wise missing} data.
For the same genomic example mentioned above, not all samples are completely observed in both data sets. Some may only have gene expression measurements, while others may only have methylation profiles.
For such a missing structure, it is very difficult to impute or integrate different data sources in a principled, unified fashion.

Integrative analysis of multi-source data has drawn more attention to the statistical learning literature lately.
Many recent approaches have been developed for integrated analysis of multi-source data \citep{tseng2015integrating}.
For example, \citet{iClusterShen2009} introduced an integrative clustering model (iCluster) which incorporates all of the data sources in a single clustering analysis.
It captures the association and shared clustering between different data sets through a joint latent variable model, but does not consider the unique aspects of each data set.
Several recent methods strive to identify not only the shared structure across multiple sources (i.e., \emph{joint}) but the structure that is specific to each source (i.e., \emph{individual}).  
\citet{Lock2013} developed the Joint and Individual Variation Explained (JIVE) method, which is an extension of the principal component analysis (PCA) to the multi-source data.
Supervised integrated factor analysis (SIFA, \citealp{li2017incorporating}) is another method which focuses on the integrative dimension reduction of multi-source data.
Several other approaches that capture joint and individual latent structures have been developed, including extensions of partial least squares \citep{Lofstedt2011}, canonical correlation analysis \citep{zhou2015}, non-parametric Bayesian modeling \citep{ray2014}, non-negative factorization \citep{yang2016}, common orthogonal basis extraction \citep{zhou2016group} and simultaneous component analysis \citep{schouteden2014}.  
However, these approaches either explicitly assume a Gaussian model or are only appropriate for continuous data. 

Batch adjustment techniques \citep{leek2010tackling,johnson2007adjusting,fan2016heterogeneity} also involve the integration of different sources of data.   They adjust raw data across different sample sets by removing batch effects caused by different laboratories or other sources of artificial heterogeneity.
However, those approaches handling batch effects are designed for Gaussian data only and can not handle block-wise missing structure.

More efforts are needed for the integrative analysis of data with different types (e.g., count and binary), as heterogeneous data are often encountered due to the disparate nature of multi-source data.
The iCluster+ approach \citep{iClustePWang2013}, which enhanced iCluster, provides a feasible approach to the clustering of multi-source data with both discrete and continuous values.
Very recently, \cite{li2017general} developed a generalized association study (GAS) framework for the multivariate association analysis of heterogeneous multi-source data.
However, none of the existing methods can easily accommodate block-wise missing values.

Block-wise missing structure is ubiquitous in multi-source data sets.
Some well known missing value imputation approaches, such as Expectation-Maximum (EM), iterative singular value decomposition (SVD), and matrix completion \citep{Mazumder2010} are effective to impute data that are missing at random in a single data set.
However, the assumption of missing at random is not valid for block-wise missing data and most existing imputation methods are not robust when the missing rate is high \citep{Xiang2014}.
The standard imputation methods are inappropriate and inefficient for block-wise missing data imputation \citep{Yuan2012}.
In many applications, a common practice to deal with block-wise missing data is to simply remove the observations with missing entries.
However, such a procedure may greatly reduce the number of observations and lead to a loss of information.
The incomplete multi-task feature learning (iMSF, \citealp{Yuan2012}) framework conducted a consistent feature selection procedure by avoiding direct block-wise missing imputation.
A bi-level learning model \citep{Xiang2014} further extended the iMSF approach to performing covariates-level and source-level analyses at the same time.
However, both methods bypass the imputation step when encountering data sets with block-wise missing entries, and thus may have limited generalizability in other contexts.
Recently, \citet{Cai2016} developed a structured matrix completion (SMC) method for imputing structured missing data using a Schur completion. 
SMC can potentially be used for block-wise missing data imputation.
However, by design, SMC is only suitable for Gaussian data, and cannot easily handle more than two data sets with heterogeneous data types.

In this paper, we develop a flexible approach for the dimension reduction of multi-source data that allows different sources to have different data types.
By assuming each data source comes from one type of distribution in the exponential family, we simultaneously model joint and individual patterns of the underlying natural parameters across data sources.
The proposed method can be applied to block-wise missing data and achieve superior imputation performance.
We devise a computationally efficient algorithm for model fitting.
We also introduce an adapted Bayesian information criterion (BIC) to select the underlying ranks of the model (i.e., the ranks of latent joint and individual structures in the model).

The rest of the paper is organized as follows.
In Section 2, the proposed models and identifiability issues are introduced for non-missing and block-wise missing data.
In Section 3, we introduce the algorithm and the rank selection.
In Section 4, we conduct comprehensive simulation studies to evaluate the performance of the proposed method and compare with existing methods.
In Section 5, we apply the proposed method to a mortality study and discuss the performance of estimation and imputation by comparison with several ad hoc methods.
\section{Generalized Integrative PCA Model}

Let $\mathbf{X}_k$ $(k = 1,\dots K)$ be an $n\times p_k$ data matrix, with $n$ being the number of samples and $p_k$ being the number of variables.
Samples are matched across $K$ data sources.
Each entry in the data matrix $\mathbf{X}_k$ is a realization of a random variable from an exponential family distribution.  
The entries of different data matrices may follow different distributions (e.g., Gaussian, Poisson, binomial), while those in the same data matrix are assumed to have the same distributional form.  
That is, each entry $x_{k,ij}$ in the $k$th data set is a realization of a random variable following a single-parameter distribution in the exponential family with an underlying natural parameter $\theta_{k,ij}$.  
The canonical form of the probability density function for each entry can be expressed as,
$$
x_{k,ij} \sim  f_k(x_{k,ij}|\theta_{k,ij})\propto \exp\{x_{k,ij}\theta_{k,ij}-b_{k}(\theta_{k,ij})\}.
$$
where $b_{k}(\theta_{k,ij})$ is a convex function which defines the distribution.
The canonical link function for the generalized linear regression is $g_{k}(\mu_{k,ij}) = b_{k}'^{-1}(\mu_{k,ij})$, where $\mu_{k,ij}$ is the mean of $x_{k,ij}$.
The entries are assumed independent given the underlying natural parameters.
We denote the underlying natural parameters matrix for $\mathbf{X}_k$ as $\boldsymbol{\Theta}_k$. 
The natural parameter matrix for all data is denoted as $\boldsymbol{\Theta}=(\boldsymbol{\Theta}_1,\dots,\boldsymbol{\Theta}_K)$, which has $p=\sum_{k=1}^Kp_k$ columns.

\subsection{Model for Non-Missing Data}
\label{nonmissing}
We first discuss our proposed model in the context of non-missing (complete) data.
For the integrated analysis of multi-source data sets, both shared and individual structure should be considered in the decomposition procedure \citep{Lock2013}.
The natural parameter matrix $\mathbf{\Theta}_k,$ for each data set, is decomposed into joint and individual latent components as follows:
\begin{eqnarray}\label{GIPCA}
&&\mathbf{\Theta}_k = \mathbf{1}\boldsymbol{\mu}_k^T+\mathbf{U}_0\mathbf{V}^T_k+\mathbf{U}_k\mathbf{A}^T_k.
\end{eqnarray}
In Model (\ref{GIPCA}), $\boldsymbol{\mu}_k$ is the column means of natural parameters and $\mathbf{1}$ is an $n\times 1$ vector of all $1$s.
Thus, natural parameters within one matrix may have different column means.
The second term $\mathbf{U}_0\mathbf{V}^T_k$ represents the shared structure among different data sources, where $\mathbf{U}_0$ is an $n\times r_J$ joint score matrix among $K$ data sets and $\mathbf{V}_k$ is a $p_k\times r_J$ joint loading matrix for $k$th data set, with $r_J\le \min(n,p_1,p_2,\dots,p_K)$ being the rank of the joint structure.
The individual structure is denoted by $\mathbf{U}_k\mathbf{A}^T_k$, where $\mathbf{U}_k$ is an $n \times r_{A_k}$ individual score matrix and $\mathbf{A}_k$ is a $p_k\times r_{A_k}$ individual loading matrix.
The individual rank for each data set is $r_{A_k}$, and $r_{A_k}\le \min(n,p_k)$.

Equivalently, the decomposition of the natural parameter matrix $\mathbf{\Theta}$ can be expressed as follows:

\begin{eqnarray}
\boldsymbol{\Theta} = (\mathbf{1},\mathbf{U}_0,\mathbf{U}_1,\dots,\mathbf{U}_K)\begin{pmatrix}
\boldsymbol{\mu}^T_1 & \dots & \boldsymbol{\mu}^T_K\\
\mathbf{V}^T_1 &  \dots & \mathbf{V}^T_K\\
\mathbf{A}^T_1 & \dots & \mathbf{0}\\
\vdots & \ddots & \vdots \\
\mathbf{0} & \dots & \mathbf{A}^T_K\\
\end{pmatrix}
 = (\mathbf{1},\mathbf{U}_0,\mathbf{U}_A)\begin{pmatrix}
 \boldsymbol{\mu}^T \\
 \mathbf{V}^T\\
 \mathbf{A}^T
 \end{pmatrix}
\label{GIPCAdecomp}
\end{eqnarray}

\noindent where $\boldsymbol{\mu}=(\boldsymbol{\mu}_1^T,\cdots,\boldsymbol{\mu}_K^T)^T$ is the concatenation of the column means for each data set, $\mathbf{V} = (\mathbf{V}^T_1,\dots,\mathbf{V}^T_K)^T$ is the concatenation of the joint loading matrices, 
$ \mathbf{U}_A = (\mathbf{U}_1,\dots,\mathbf{U}_K)$ is the concatenation of the individual score matrices, $\mathbf{A} = diag(\mathbf{A}_1,\dots,\mathbf{A}_K)$ is a block-wise diagonal matrix, and $\mathbf{0}$ represents any zero matrix with compatible size.

In particular, when there is only one data set, our proposed GIPCA reduces to the decomposition of one natural parameter matrix, which coincides with the exponential family principal component analysis (EPCA, \citealp{collins2001generalization}).
Under the Gaussian assumption with equal variance, the decomposition of natural parameter matrix reduces to a factorization of the original multi-source data set.
Thus, our Model (\ref{GIPCAdecomp}) is identical to JIVE \citep{Lock2013} in this context.
With just two data sets, Model (\ref{GIPCAdecomp}) coincides with the GAS model \citep{li2017general} applied to data sets without missing values.

\subsection{Model with Block-wise Missing Data}
\begin{figure}[!ht]
\caption{GIPCA for block-wise missing data}
\caption*{The three big rectangles represent three data sets $\mathbf{X}_k$ $(k = 1,2,3)$ with block-wise missing values (i.e., blank strips).
The horizontal direction (rows) represents samples in the three data sets.
And the vertical direction (columns) represents variables.
The grey color in the big rectangles means that the data are observed for the corresponding samples in the corresponding sources.
The blank rectangles are the block-wise missing entries.
Those rectangles on the side are joint score and loading matrices $\mathbf{U}_0$, $\mathbf{V}_k$ and individual score and loading matrices $\mathbf{U}^*_k$, $\mathbf{A}_k$.}
\label{fig:GEPCA}
\includegraphics[width = 1\linewidth]{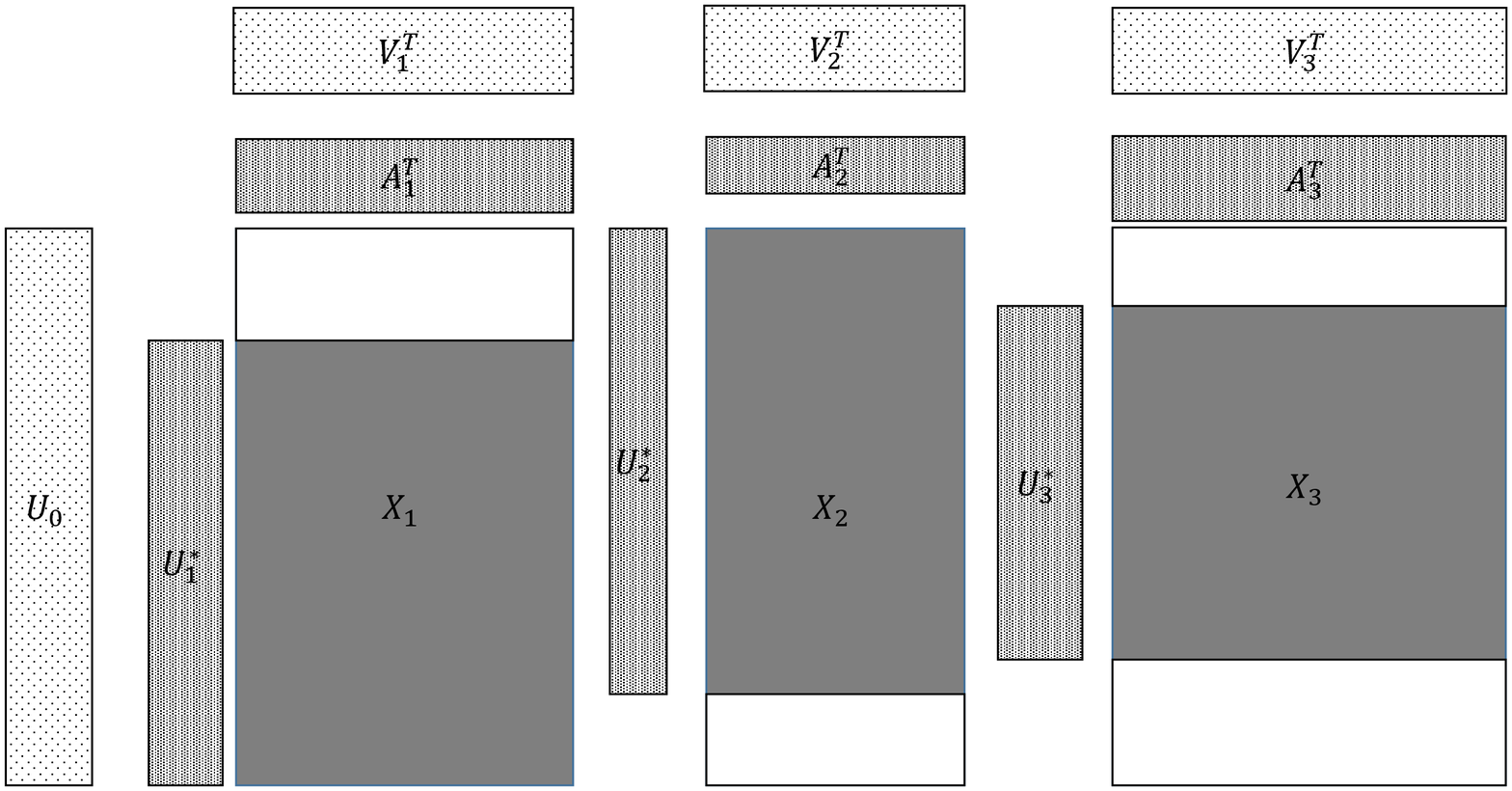}
\end{figure}
We extend the model described in Section~\ref{nonmissing} to allow for block-wise missing structure.
Figure \ref{fig:GEPCA} is an illustrative picture of data sets with block-wise missing.
Due to the block-wise missing entries in the data sets, the corresponding rows in the individual score matrices $\mathbf{U}_k$ in Model (\ref{GIPCAdecomp}) are missing.
Thus we denote the submatrix of $\mathbf{U}_k$ containing only rows without block-wise missing as $\mathbf{U}^*_k$.
The joint score matrix $\mathbf{U}_0$ re
s the same as in Model (\ref{GIPCAdecomp}) because for all samples at least one data source is with complete observations, which helps us identify joint structure.

With block-wise missing data, for each data set $k$, the decomposition of the natural parameter matrix underlying the observed data becomes
\begin{eqnarray}\label{model:blockGIPCA}
\boldsymbol{\Theta}^*_k = \mathbf{1}\boldsymbol{\mu}_k^T+\mathbf{U}_0^{[k]}\mathbf{V}^T_k+\mathbf{U}^*_k\mathbf{A}^T_k,
\end{eqnarray}
where $\boldsymbol{\Theta}^*_k$ is an $n_k\times p_k$ matrix (a submatrix of $\boldsymbol{\Theta}_k$ in Model \eqref{GIPCA}).
The joint score matrix $\mathbf{U}_0^{[k]}$ is an $n_k\times r_J$ submatrix of $\mathbf{U}_0$, where only the rows corresponding to the complete samples in the $k$th data source are kept.
The individual score matrix $\mathbf{U}^*_k$ is an $n_k\times r_{A_k}$ matrix. 
The means $\boldsymbol{\mu}_k$, the joint and individual loading matrices $\mathbf{V}_k$, $\mathbf{A}_k$ remain the same as in Model (\ref{GIPCAdecomp}).
We also note that $\mathbf{1}$ is an $n_k\times 1$ vector of all $1$s.
When there is no missing value, Model \eqref{model:blockGIPCA} exactly coincides with Model \eqref{GIPCAdecomp}.

We remark that despite the block-wise missingness, the joint structure in Model (\ref{model:blockGIPCA}) across data sources is $\mathbf{U}_0\mathbf{V}$. 
For a sample with block-wise missing values, as long as it has observations in some data sources, it provides information towards the shared structure.
Thus, the underlying joint score matrix is complete, regardless of the block-wise missing structure.
The mechanism of block-wise missing imputation relies on the joint structure.
Such shared information among different data sets informs the missing data for each data source.
Specifically, once estimated, the means and the joint structure can be effectively used to impute block-wise missing data.

\subsection{Identifiability Conditions}\label{id}
In order to ensure identifiability of the estimation, the model parameters should satisfy certain conditions. Following the discussion in \citep{Lock2013,li2017general}, we provide the identifiability conditions for Model (\ref{model:blockGIPCA}) as the following.
\begin{enumerate}[noitemsep,topsep=0.5pt]
\item The columns of the score matrices $\mathbf{U}_0$, $\mathbf{U}_k$ are linearly independent and the columns of the means $\boldsymbol{\mu}_k$ and the loading matrices $\mathbf{V}$, $\mathbf{A}_k$ within each data set are linearly independent.
\item All the score matrices are column-centered and the column space of the joint score matrix is orthogonal to the column space of the individual score matrices. 
\item All the separate score and loading matrices have orthogonal columns. 
\end{enumerate}
The first condition ensures the joint and individual structures are clearly separable.
The second orthogonality condition enhances the interpretability by requiring that the means, joint and individual structures are orthogonal to each other. 
The third condition rules out arbitrary rotations within each subspace. 
The above conditions guarantee that the model is fully identifiable (up to some trivial order switch and scale change).

\section{Algorithm}
\label{alg}

In this section, we explain how we estimate each parameter in Model (\ref{model:blockGIPCA}).
We first assume the ranks for the shared and individual structures are known and devise an iterative algorithm for model fitting.
Then we introduce an adaptive BIC procedure for rank selection, which is tailored for the proposed approach. 

\subsection{GIPCA algorithm}
The unknown parameters in Model (\ref{model:blockGIPCA}) are estimated by maximizing the joint log likelihood.
Under the assumption that individual measurements are mutually independent given the underlying natural parameter matrix, the maximum likelihood estimators (MLE) are
\begin{eqnarray}
(\hat{\boldsymbol{\mu}},\hat{\mathbf{U}}_0,~\hat{\mathbf{U}}^*_1,\cdots,\hat{\mathbf{U}}^*_K,~\hat{\mathbf{V}},~\hat{\mathbf{A}}) = \argmax _{\substack{\mathbf{\Psi}}} \sum_{k = 1}^K \sum_{i = 1}^{n_k} \sum_{j = 1}^{p_k} \log f_{k}(x_{k,ij}|\theta_{k,ij})
\label{model:objectivefunc}
\end{eqnarray}
where $\mathbf{\Psi} = \{\boldsymbol{\mu}_k,\mathbf{U}_0,\mathbf{V}_k,\mathbf{U}^*_k,\mathbf{A}_k,~k = 1,\dots,K\}$ is the set of unknown parameters, $\boldsymbol{\Theta}^*_k = (\theta_{k,ij})$ has the decomposition structure in Model (\ref{model:blockGIPCA}) and $f_k(\cdot)$ is the probability density function in each data source.

It is computationally prohibitive to directly maximize the log likelihood because the objective function is not convex with respect to all the parameters.
As a remedy, we exploit a block coordinate descent algorithm to estimate the parameters.
Namely, we alternatively estimate the joint structure along with the intercept and the individual structures until converge.
More specifically, we
\begin{itemize}[noitemsep,topsep=0.5pt]
\item fix $\mathbf{U}^*_k$ and $\mathbf{A}_k$ for all data sets, and estimate $\boldsymbol{\mu}_k$,  $\mathbf{U}_0$ and  $\mathbf{V}$;
\item fix $\boldsymbol{\mu}_k$, $\mathbf{U}_0$ and $\mathbf{V}_k$, and estimate $\mathbf{U}^*_k$ and $\mathbf{A}_k$ in each data set.
\end{itemize}
Consequently, the computation is significantly simplified. We shall provide more details below.

We first estimate the intercept and joint structure with the individual structures fixed.
To further alleviate the computational burden, we fix the joint score matrix $\mathbf{U}_0$ to estimate the joint loading matrix $\mathbf{V}=(\mathbf{V}_1^T,\cdots,\mathbf{V}_K^T)^T$ and the intercept $\boldsymbol{\mu} = (\boldsymbol{\mu}^T_1,\dots,\boldsymbol{\mu}^T_K)^T$.
The estimation of each row in $\mathbf{V}$ paired with the corresponding entry in $\boldsymbol{\mu}$ can be cast as a generalized linear model (GLM) estimation problem.
More specifically, let $\boldsymbol{\theta}^*_{k\cdot j}$ be $j$th column of $\boldsymbol{\Theta}^*_k$.
We have
$\boldsymbol{\theta}^*_{k\cdot j} = \mathbf{1}^T\mu_{kj}+\mathbf{U}_{0}^{[k]}\mathbf{v}^T_{k\cdot j}+\mathbf{U}^*_{k}\mathbf{a}^T_{k\cdot j}$,
where $\mu_{kj}$ is the $j$th entry of $\boldsymbol{\mu}_k$, $\mathbf{v}_{k\cdot j}$ is the $j$th row of $\mathbf{V}_k$,  and $\mathbf{a}_{k\cdot j}$ is the $j$th row of $\mathbf{A}_k$.
The estimation of $\mathbf{v}_{k\cdot j}$ and $\mu_{kj}$ can be obtained by fitting a GLM with the canonical link function, and $\mathbf{U}^*_{k}\mathbf{a}^T_{k\cdot j}$ being the offset.
Similarly, when we fix the joint loading matrix $\mathbf{V}$ to estimate the joint score matrix $\mathbf{U}_0$, again this can be formulated as a GLM problem.
Let $\boldsymbol{\theta}_{i \cdot} = (\boldsymbol{\theta}_{1i \cdot}^T,\dots,\boldsymbol{\theta}_{Ki \cdot}^T)^T$, where $\boldsymbol{\theta}_{ki \cdot}$ is the column vector of $i$th row of $\boldsymbol{\Theta}_k$.
We have
$\boldsymbol{\theta}_{i \cdot} = (\boldsymbol{\mu}_1^T,\dots,\boldsymbol{\mu}_K^T)^T+\mathbf{V}\mathbf{u}_{0i \cdot}+(\mathbf{A}_1\mathbf{u}_{1i \cdot},\dots,\mathbf{A}_K\mathbf{u}_{Ki \cdot})$,
where $\mathbf{u}_{0i \cdot}$ is the column vector of the $i$th row in joint score matrix $\mathbf{U}_{0}$ and $\mathbf{u}_{ki \cdot}$ is the column vector of the $i$th row in individual score matrix $\mathbf{U}_k$.
We remark that the standard GLM model fitting procedure cannot be directly applied to the estimation of $\mathbf{u}_{0i \cdot}$ since the canonical link functions are different for different data types across multiple sources.
To address this, we follow \citet{li2017general} and adopt an iteratively reweighted least squares algorithm (IRLS, \citealp{mccullagh1989generalized}) to accommodate heterogeneous link functions.

Then we estimate individual structures with fixed $\boldsymbol{\mu}_k$ and the joint score $\mathbf{U}_0$ and loading $\mathbf{V}$.
Based on Model (\ref{model:blockGIPCA}), the estimation of individual structures is directly separable for each data source.
We still exploit the alternating algorithm to estimate $\mathbf{U}^*_k$ and $\mathbf{A}_k$.
Similar to the estimation of the joint structure, the estimation of $\mathbf{U}^*_k$ can be parallelized as $n_k$ GLMs, and the estimation of $\mathbf{A}_k$ can be parallelized as $p_k$ GLMs.
The estimated parameters are then plugged into the log likelihood in Model (\ref{model:objectivefunc}).

The estimates in $\hat{\mathbf{\Psi}} = \{\hat{\boldsymbol{\mu}}_k,\hat{\mathbf{U}}_0,~\hat{\mathbf{U}}^*_k,~\hat{\mathbf{V}}_, ~\hat{\mathbf{A}}_k\}$ in each iteration may not meet the identifiability conditions in Section \ref{id}.
Some regularization procedure is desired so that the conditions are satisfied and the likelihood values are unchanged. 
In order to achieve that, after each iteration, we transform the estimated parameters $\hat{\mathbf{\Psi}}$ as follows.
Define the projection matrix of the column space of $(\mathbf{1},\mathbf{U}_0)$ as $\mathbf{P}_J$.
We want to project the individual score matrices to the orthogonal complement of the column space of $(\mathbf{1},\mathbf{U}_0)$.
However,  the individual score matrix $\mathbf{U}^*_k$ does not have the same dimension as the column space of $(\mathbf{1},\mathbf{U}_0)$.
To address this, we define a new estimated individual score matrix $\hat{\mathbf{U}}_k: n\times p_k$ based on $\mathbf{U}_k^*$, where the missing observations are filled with $0$.
Then we get the projected individual score matrix $(\mathbf{I}-\mathbf{P}_J)\hat{\mathbf{U}}_k$.
Column-center the submatrix of the projected individual score matrix containing only complete samples and denote it as  $\check{\mathbf{U}}_k: n_k\times p_k$.
Apply SVD to the new individual structures $\check{\mathbf{U}}_k\hat{\mathbf{V}}_k^T$, let the left singular vectors absorb the singular values.
Let a score matrix $\check{\mathbf{U}}^*_k: n\times p_k$ be based on the left singular vector and corresponding block-wise missing rows filled $0$.
The new joint structure after identifiability modification is the concatenation of $K$ matrices where each is  
$\mathbf{1}\hat{\boldsymbol{\mu}}_k+\hat{\mathbf{U}}_0\hat{\mathbf{V}}_k^T+\hat{\mathbf{U}}_k\hat{\mathbf{A}}_k^T-\check{\mathbf{U}}^*_k\check{\mathbf{V}}_k^{*T}$.
The column mean of each new joint structure is $\check{\boldsymbol{\mu}}_k^*$.
Apply SVD to the concatenation of each column-centered joint structure, and let the left singular vectors absorb the singular values.
We denote the new score and loading matrices as $\check{\mathbf{U}}_0^*,~\check{\mathbf{V}}^* $.
Consequently, the modified estimators $\check{\boldsymbol{\mu}}^*_k,\check{\mathbf{U}}^*_0,~\check{\mathbf{U}}^*_k,~\check{\mathbf{V}}^*, \check{\mathbf{A}}^*_k$ satisfy all the conditions.

The iterative algorithm terminates when the difference of the log likelihood between the previous step and current step is smaller than a prefixed threshold.
Our proposed algorithm is a block coordinate descent algorithm which ensures the log likelihood in each step of the algorithm is non-decreasing.
Thus, the algorithm is guaranteed to converge.
We summarize the model fitting algorithm with known ranks for our proposed method in Algorithm \ref{alg:GIPCA}.

\begin{algorithm}
\caption{GIPCA algorithm}\label{alg:GIPCA}
\begin{algorithmic}[H]
\State Set initial values of each element in Model (\ref{model:blockGIPCA}) as $\mathbf{\Psi}^{(0)}$
\While{The convergence criterion does not satisfy}
\State In the $l$th iteration:
    \State Fix the individual structure: $\mathbf{U}_k^{*(l-1)}$,$\mathbf{V}_k^{(l-1)},~k = 1,2,\dots,K$,
        \State\hspace{\algorithmicindent} Fix $\mathbf{U}_0^{(l-1)}$, estimate each row of $(\boldsymbol{\mu}_k,\mathbf{V}_k)$ via GLM.
        \State\hspace{\algorithmicindent} Fix $\mathbf{V}^{(l)}$ and $\boldsymbol{\mu}^{(l)}$, estimate each row of $\mathbf{U}_0$ by the adapted IRLS algorithm.
    \State Fix the means and joint structure $(\boldsymbol{\mu}^{(l)},\mathbf{U}_0^{(l)},\mathbf{V}^{(l)})$
       \State\hspace{\algorithmicindent} Fix $\mathbf{U}_k^{*(l-1)}$, estimate each row of $\mathbf{V}_k$ via GLM.
       \State\hspace{\algorithmicindent} Fix $\mathbf{V}_k^{(l)}$, estimate each row of $\mathbf{U}_k^{*}$ via GLM.
    \State Conduct the regularization procedure to satisfy the identifiability conditions.
    \State Plug the estimated parameters to the log likelihood function.
\EndWhile
\end{algorithmic}
\end{algorithm}

After obtaining the estimates of the parameters in $\mathbf{\Psi}$, we impute the block-wise missing entries using the shared parameters.
More specifically, we use the same procedure as the regularization step mentioned above to get $\hat{\mathbf{U}}_k: n\times r_{A_k}$.
Then, we can have the estimated complete natural parameter matrix,
$\hat{\mathbf{\Theta}}_k = \mathbf{1}^T\hat{\boldsymbol{\mu}}_k+\hat{\mathbf{U}}_0\hat{\mathbf{V}}_k+\hat{\mathbf{U}}_k\hat{\mathbf{A}}_k$.
In particular, the estimated natural parameter matrix for block-wise missing entries is, $\hat{\mathbf{\Theta}^\dagger}_k = \mathbf{1}^T\hat{\boldsymbol{\mu}}_k+\hat{\mathbf{U}}_0^{[k]^c}\hat{\mathbf{V}}_k$, where $\hat{\mathbf{U}}_0^{[k]^c}: (n-n_k)\times p_k$ is the complement submatrix of $\hat{\mathbf{U}}_0^{[k]}$ in terms of $\hat{\mathbf{U}}_0$.
Each vector $\mathbf{1}$ is with the compatible size.
By taking the inverse of the link function, the imputed data is $g^{-1}_k(\hat{\mathbf{\Theta}^\dagger}_k )$.

\subsection{Rank Estimation: BIC}
\label{sec:BIC}
There are many approaches in the PCA literature to determine the number of principal components or the rank of the latent structure.
For example, one may exploit scree plots of eigenvalues to choose the rank that explains a certain proportion of the total variation \citep{Jolliffe1986PCA}, or use a hypothesis testing procedure (e.g., Bartlett's test) to determine the rank.
There is a large amount of literature on selecting ranks for matrix decomposition with Gaussian assumption.
However, there is only a little considering rank estimation for non-Gaussian data.
 \citet{LandgrafAJ2015} proposed an approach for binary data based on the percentage of deviance explained by some principal components.
As to the rank selection for multi-source data, a permutation testing approach was proposed to JIVE \citep{Lock2013}.
BIC is another approach that is adapted to JIVE to implement rank selection \citep{o2016r}.
A two-step cross-validation method \citep{li2017general} used the sum of squared Pearson residuals as the criterion to select ranks when modeling heterogeneous data with exponential family distributions assumption.
Nevertheless, none of the literature mentioned rank estimation for a multi-source data with block-wise missing entries.

Here, we develop an adapted BIC approach to estimate the joint and individual ranks of the underlying natural parameter matrices for multi-source data.
The key to the adapted BIC criterion is to calculate the number of parameters to be estimated in the model.
The joint score matrix $\mathbf{U}_0$ has $\sum_{j = n-r_J}^{n-1}j$ entries to estimate since it has centered columns which are orthogonal of each other.
For each data set, there are $p_k$ unknown means in Model (\ref{model:blockGIPCA}).
Similarly for the joint and individual loading matrices, they have $\sum_{j' = p-r_J}^{p-1} j'$ and $\sum_{j'' = p_k-r_{A_k}}^{p_k-1} j''$ parameters to estimate.
The individual score matrix $\mathbf{U}_k^*$ is required to be orthogonal to the joint score matrix and the columns of individual score matrix are centralized and linearly independent of each other.
Thus the number of free parameter in the individual score matrix is $\sum_{l = n_k-(r_J+r_{A_k})}^{n_k-(r_J+1)}l$.
The number of free parameters in the data sets is,
 \begin{align}
\mathcal{K} =\sum_{k = 1}^{K}p_k+\sum_{j = n-r_J}^{n-1}j+\sum_{j' = p-r_J}^{p-1} j'+\sum_{k = 1}^K\sum_{l = n_k-(r_J+r_{A_k})}^{n_k-(r_J+1)}l+\sum_{k = 1}^{K}\sum_{j'' = p_k-r_{A_k}}^{p_k-1} j''\nonumber
 \end{align}
The number observations in data set $\mathbf{X}_k$ is $n_kp_k$.
If there is no block-wise missing in the data sets, $n_1 = n_2 = \dots,=n_K = n$.
For each combination of $r_J,r_k,~k = 1,2,\dots,K$, a BIC score could be calculated.
The value of BIC is calculated as,
\begin{eqnarray}\label{blockBIC}
\text{BIC} = -2 l^*(\mathbf{X}|\hat{\mathbf{\Psi}})+\log(\sum_{k = 1}^{k}n_k p_k)\mathcal{K}
\end{eqnarray}
where $l^*(\mathbf{X}|\hat{\mathbf{\Psi}})$ is the value of log likelihood given $\hat{\mathbf{\Psi}}$.

In practice, we use a stepwise selection approach to select the ranks via BIC.  We first compute BIC for the null model $r_J = 0,r_k = 0,k = 1,2,\dots,K$.
We add one to or deduct one from each of the ranks at a time and choose the next rank combination with the smallest BIC value.
For instance, assume we have two data sets and start from the BIC score for $r_J = 0,r_k = 0,~k = 1,2$.
Next, we calculate the BIC value for $r_J = 1~r_1 = r_2 = 0$, $r_J = r_2 =0~r_1 = 1$ and $r_J = r_1 =0~r_2 = 1$ and choose the ranks with smallest BIC score.
The selection procedure is terminated when the BIC score reaches a local minimum.
Then the ranks combination when the procedure is stopped is the estimated ranks.

\section{Simulation}
In this section, we conduct comprehensive simulation studies to validate the proposed method.
Since there is no existing method that directly addresses the multi-source multi-type data imputation problem, we come up with two ad hoc approaches to compare with our method.
\begin{itemize}[noitemsep,topsep=0.5pt]
\item []\textbf{Ad Hoc 1 (EPCA-PCA)}: First we estimate a low-rank approximation to the natural parameter matrix of each data set via EPCA.
Then, we apply PCA to the concatenated approximations across different data sources.
\item[] \textbf{Ad Hoc 2 (EPCA-SMC)}: EPCA is first applied to each data set.
Then, structured matrix completion (SMC, \citealp{Cai2016}) is applied to the estimated natural parameter matrices to impute the block-wise missing entries.
\end{itemize}

The data sets are generated from Model (\ref{GIPCAdecomp}) and we apply three different methods to the data and impute the block-wise missing entries.
For both \textbf{Ad Hoc 1} and \textbf{Ad Hoc 2} methods, if the Gaussian assumption is satisfied for some data sets, then EPCA step is ignored for such data sets and the original data sets are used in the next step (PCA or SMC).
The application of SMC is limited to two data sets when only one of them has block-wise missing entries.
Therefore, when we apply SMC to more than one data set has block-wise missing entries, we proceed with one data source at a time. 
For example, if we have two data sources where both have missing observations, apply SMC approach twice to do imputation for both data sets.

\subsection{Settings}
We set the sample size to be $n=200$ and the number of variables in each data source to be $p_k = 150$.
The joint and individual ranks for the natural parameter matrices are $r_0 = r_1 = r_2 = 2$.
Joint and individual score matrices $(\mathbf{U}_0,\mathbf{U}_1,\mathbf{U}_2)$ are filled with uniform random numbers $Unif(-0.5,0.5)$ and normalized to have orthonormal columns.
In \textbf{Scenario 4}, we try $3$ data sets with similar settings as the other scenarios.
We generate different singular values of joint structure and each individual structure for different scenarios, and the singular values are absorbed by the score matrices.

\begin{itemize}[noitemsep,topsep=0.5pt]
\item[] \textbf{Scenario 1}: \textbf{Gaussian-Gaussian}
The individual loading matrices $\boldsymbol{A}_1,~\boldsymbol{A}_2$ are filled with $Unif(-0.5,0.5)$ and normalized to have orthonormal columns.
The joint loading matrix $(\boldsymbol{V}^T_1,~\boldsymbol{V}^T_2)^T$ is generated similarly to have orthonormal columns and is projected to the complement of the column space for the individual loading matrices $diag(\boldsymbol{A}^T_1,~\boldsymbol{A}^T_2)^T$.
The singular values of the joint structure were set to be $(250,150)$, the singular values of the individual structures to be $(150,100)$ and $(150,140)$.

\item[] \textbf{Scenario 2}: \textbf{Gaussian-Poisson}
The procedure to generate individual loading matrices is similar to \textbf{Scenario 1}.
The joint loading matrices $\boldsymbol{V}_1$ for Gaussian and $\boldsymbol{V}_2$ for Poisson are generated from $Unif(-1,1)$, and $Unif(-0.25,0.25)$ respectively.
The singular values of the joint structure are set to be $(240,220)$ for joint, the singular values of the individual structures to be $(90,80)$ for Gaussian and $(90,80)$ for Poisson.

\item[] \textbf{Scenario 3}: \textbf{Gaussian-binomial}
The procedure to generate individual loading matrices is similar to \textbf{Scenario 1}.
The joint loading matrices $\boldsymbol{V}_1$ for Gaussian, $\boldsymbol{V}_2$ for binomial are generated from $Unif(-0.5,0.5)$, and $Unif(-1.5,1.5)$ respectively.
The singular values of the joint structure are set to be $(240,220)$ for joint, and the singular values of the individual structures to be $(90,80)$ for Gaussian and $(100,80)$ for binomial.

\item[] \textbf{Scenario 4}: \textbf{Gaussian-Poisson-binomial}
The joint loading matrices $\boldsymbol{V}_1$ for Gaussian, $\boldsymbol{V}_2$ for Poisson are generated from $Unif(-0.5,0.5)$, and $\boldsymbol{V}_3$ for binomial is generated from $Unif(-1.5,1.5)$.
The individual loading matrices $\boldsymbol{A}_1$ (Gaussian), $\boldsymbol{A}_2$ (Poisson), $\boldsymbol{A}_3$ (binomial) are generated from $Unif(-0.5,0.5)$, $Unif(-0.25,0.25)$, and $Unif(-1.5,1.5)$ correspondingly.
The singular values of the joint structure are set to be $(300,280)$, the singular values of the individual structures to be $(150,120)$ for Gaussian, $(150,140)$ for Poisson and $(200,180)$ for binomial.

\item[] \textbf{Scenario 5}: \textbf{ Poisson-binomial}
The joint loading matrices $\boldsymbol{V}_1$ for binomial, $\boldsymbol{V}_2$ for Poisson are generated from $Unif(-1.5,1.5)$, and $Unif(-0.5,0.5)$ respectively.
\end{itemize}
The means for Gaussian data set in each scenario that contains Gaussian data are generated from $Unif(-0.5,0.5)$.
For Poisson distribution, the inverse of the canonical function makes the realizations skewed to $0$ if the natural parameter is a negative number with large absolute value and skewed to a large positive number if the natural parameter is a large positive number.
Thus, the scale of the natural parameter matrix for Poisson distribution is required to be smaller in \textbf{Scenarios 2, 4, 5}.
We also set the means of Poisson distribution to be positive (from $Unif(0,1)$) to mimic Poisson data in reality.
For binomial distribution, we increase the singular values to boost the signal level of binomial data in \textbf{Scenarios 3, 4, 5}.
The means for binomial data set are generated from $Unif(-1.5,1.5)$.

When the natural parameters are fixed, data are generated from the corresponding distributions.
For Gaussian data, we set the variance for the generated data to be $1$.
For binomial data, we set the number of trials to be $100$.
For each simulation, we randomly pick some rows in each data set to be missing.
Those rows should not be overlapped over all the data sets to ensure that for each sample, data from at least one data set are without missing.
Different missing rates  ($5\%$ or $10\%$ for rank selection and $5\%$ or $15\%$ for missing imputation) are applied to the generated data when we compare our proposed method with other existing methods.
We repeat the procedure multiple times to evaluate the rank selection performance and compare the imputation accuracy of different methods.

\subsection{Result}
\label{SimResult}
When the natural parameter matrix for each scenario is fixed, we apply the rank selection procedure mentioned in Section \ref{sec:BIC} to the data generated from corresponding distribution independently for $50$ times.
BIC criterion (Model \eqref{blockBIC}) is used to estimate ranks for each simulation scenarios with different missing rates.
We apply the proposed BIC criterion to all the scenarios with different missing rate $0\%,5\%$, $10\%$.
The results of rank estimation with different simulation scenarios and missing rates are shown in Table S1 in the supplementary materials.

Overall, the adapted BIC criterion performs well for different settings.
The stepwise selection procedure correctly identifies the true ranks for joint structure and individual structures almost all the times for scenarios with two data types with various missing rates.
We also apply the selection procedure to \textbf{Scenario 4} with three data types: Gaussian, Poisson, and binomial.
However, for this scenario the BIC-selected ranks tend to be close to the truth but misallocated;  the majority of the 50 simulations select the joint rank to be $3$, individual ranks to be $1$ for Gaussian and Poisson, and 2 for binomial. 
This may be because the signal-to-noise ratio for the binomial data is relatively low compared to the other datasets.  Alternative approaches to rank selection that can accommodate to multiple ($>$2) sources of data call for more investigation.

When the natural parameter matrix is fixed, we generate data from corresponding distributions independently for $100$ replications.
We compare the two ad hoc methods and our proposed method by applying them to the simulated data to estimate elements of $\mathbf{\Psi}$ in Model (\ref{model:blockGIPCA}).
We evaluate the imputation accuracy by the relative Frobenius loss.
Mathematically, the relative Frobenius loss is defined as, 
\begin{align}\label{LossMiss}
\text{DiffR}_{Miss} = \frac{||\boldsymbol{\Theta}_k^\dagger-\hat{\boldsymbol\Theta}_k^\dagger||_{F}}{||\boldsymbol\Theta_k^\dagger||_{F}}
\end{align}
where $\mathbf{\Theta}_k^\dagger$ and $\hat{\mathbf{\Theta}}^\dagger_k$ are true and estimated natural parameter matrices for block-wise missing entries.

\begin{table}[!ht]
\setlength\tabcolsep{2pt}
\tablinesep=1ex
\centering
\caption{Simulation results for two data sets based on $100$ simulation runs when the natural parameter matrices are fixed for each data source.
5\%M, 15\%M represent the 5\% and 15\% missing rate correspondingly.
The median and the median absolute deviation (MAD) for the relative Frobenius loss under each scenario are calculated.
MAD is in parenthesis.
The best results are highlighted in bold.}
\small

\begin{tabular}{ccllllll}
\hline\hline
 &\multicolumn{2}{c}{\textbf{Adhoc1}} &\multicolumn{2}{c}{\textbf{Adhoc2}}  & \multicolumn{2}{c}{\textbf{GIPCA}}\\
 && Source1 &Source2 &Source1 &Source2&Source1 &Source2\\
\hline

\textbf{Scenario 1 (5\%M)}& $DiffR_{Miss}$
&  8.46 (2.21) &8.78 (2.07)  &1.13(0.15) &1.00(0.00)& \textbf{0.69 (0.01)} &\textbf{0.69 (0.00)}\\
\textbf{Gaussian Gaussian } & \color{blue}\textbf{Running Time} &\multicolumn{2}{c}{\color{blue}98.12 (12.06)}& \multicolumn{2}{c}{\color{blue}3.25 (0.04)} &\multicolumn{2}{c}{\color{blue}96.66 (21.79)}\\

\textbf{Scenario 1 (15\%M)}& $DiffR_{Miss}$
& 7.44 (0.45) & 7.64 (0.46) & 1.36 (0.00) &1.01 (0.00) &  \textbf{0.65 (0.00)} & \textbf{0.72 (0.00)}\\
\textbf{Gaussian Gaussian } & \color{blue}\textbf{Running Time} &\multicolumn{2}{c}{\color{blue}98.12 (12.06)}& \multicolumn{2}{c}{ \color{blue}3.25 (0.04)} &\multicolumn{2}{c}{\color{blue}96.66 (21.79)}\\

\textbf{Scenario 2 (5\%M)}&$DiffR_{Miss}$
& 8.13 (5.20) & 2.87 (1.00) & 0.49 (0.01) & 1.31 (0.88) &\textbf{0.45 (0.00)}&  \textbf{0.28 (0.01)}\\
\textbf{Gaussian Poisson } & \color{blue}\textbf{Running Time} & \multicolumn{2}{c}{238.69 (53.45)} &\multicolumn{2}{c}{2.23 (0.05)} & \multicolumn{2}{c}{209.77 (65.14)}\\

\textbf{Scenario 2 (15\%M)}&
$DiffR_{Miss}$
& 9.40 (3.01) & 3.64 (0.29) & 0.72 (0.08) & 0.58 (0.03) & \textbf{0.46 (0.00)} &\textbf{0.47 (0.00)} \\
\textbf{Gaussian Poisson } & \color{blue}\textbf{Running Time} & \multicolumn{2}{c}{\color{blue}245.42 (53.78)} & \multicolumn{2}{c}{\color{blue}1.84 (0.05)} & \multicolumn{2}{c}{ \color{blue}221 (54.84)}\\

\textbf{Scenario 3 (5\%M)}&
$DiffR_{Miss}$
&  1.11 (0.40) & 1.02 (0.32) & 0.84 (0.00) &0.99 (0.00) &\textbf{0.77 (0.00)}& \textbf{0.43 (0.00)}\\
\textbf{Gaussian binomial } &\color{blue}\textbf{Running Time} &\multicolumn{2}{c}{ \color{blue}464.97 (93.89)} &\multicolumn{2}{c}{\color{blue}1.82(0.11)}& \multicolumn{2}{c}{\color{blue}193.94 (110.07)}\\

\textbf{Scenario 3 (15\%M)}&
$DiffR_{Miss}$ & 1.11 (0.40) & 1.02 (0.32) & 0.84 (0.00) &0.99 (0.00) & \textbf{0.77 (0.00)} & \textbf{0.43 (0.00)}\\
\textbf{Gaussian binomial } &\color{blue}\textbf{Running Time} & \multicolumn{2}{c}{\color{blue}473.5 (99.23)} & \multicolumn{2}{c}{\color{blue}1.83 (0.11)} & \multicolumn{2}{c}{\color{blue}189.65 (106.73)}\\

\textbf{Scenario 5 (5\%M)}&
$DiffR_{Miss}$& 6.1 (2.07) &  4.84 (0.88) & 0.58 (0.03) & 4.58 (1.10) & \textbf{0.57 (0.01)} & \textbf{0.84 (0.00)} \\
\textbf{Poisson binomial } &\color{blue}\color{blue}\textbf{Running Time}& \multicolumn{2}{c}{\color{blue}188.99 (27.49)} & \multicolumn{2}{c}{\color{blue}0.38 (0.03)} & \multicolumn{2}{c}{\color{blue}914.98 (177.24)}\\

\textbf{Scenario 5 (15\%M)}&
$DiffR_{Miss}$& 5.66 (1.66) & 5.35 (1.36) & 0.63 (0.02) & 1.47 (0.21) &  \textbf{0.61 (0.01)} & \textbf{0.86 (0.00)}\\
\textbf{Poisson binomial } &\color{blue}\textbf{Running Time}& \multicolumn{2}{c}{\color{blue}191.19 (33.53)} & \multicolumn{2}{c}{\color{blue}0.30 (0.01)}&\multicolumn{2}{c}{\color{blue}808.15 (210.31)}\\
\hline\hline
\end{tabular}
\label{SimCompBMiss}
\end{table}

The simulation results for two data sets are shown in Table \ref{SimCompBMiss} and for three data sets are shown in Table S2 in the supplementary materials.
For all scenarios, the imputation accuracy of GIPCA outperforms the other ad hoc methods with different distribution combinations and different missing rates.
Neither EPCA-PCA nor SMC considers partitioning the joint association from individual structure.
We also check the Frobenius norm for the difference between the estimated and true means, the relative Frobenius loss for natural parameter matrix without missing entries.
We note that under \textbf{Scenario 1}, both EPCA-PCA and GIPCA have similar performance.
Under the Gaussian assumption, EPCA-PCA reduces to PCA with the sum of ranks and GIPCA reduces to JIVE without missing entry (Section \ref{nonmissing}).
Therefore, their estimation accuracy to estimate natural parameter matrix corresponding to samples without block-wise missing is close to each other.

In addition to the simulation settings above, we also explore the scenarios when the signals of the joint and individual structures are distinct.
We set the true singular values of the natural parameter matrix of the joint structure relatively small ($1/2$, $1/5$ or $1/10$ of the singular values in the original setting in different scenarios).
The results are shown in Table S3 in the supplementary materials.
The results show that the performance of missing imputation for \textbf{Gaussian-Gaussian} and \textbf{Gaussian-Poisson} scenarios is relatively robust against the change of singular values.
For scenarios involving binomial distributions, the performance is sensitive to the change of signal.

In order to evaluate how sensitive the algorithm is to initial values, we use different initial values and evaluate the estimation performance.
Data are generated in the same way in Section 4.1.
For each scenario, we fix the simulated data and generate different initial values based on different random seeds.
Table S4 in the supplementary materials shows that the performance of missing imputation by the proposed method is stable, which indicates that our algorithm is not sensitive to different initial values.

\section{Real Data Analysis}

In this section, we apply our proposed method to a mortality study, where the data are publicly available from Human Mortality Database \citep{HMD}.
We focus on exposure-to-risk and population size data sets in two countries, Italy and Switzerland, and analyze the commonality and specificity of the mortality rate patterns in both countries.
The chosen data set, exposure-to-risk data set contains realizations of binomial random variables with the number of trials equal to the corresponding entries in population size data set.
The Italian data have $143$ rows where each row represents a year between $1872-2014$; the Switzerland data have $139$ rows where each row represents a year between $1876-2014$.
Since the number of exposure-to-risk becomes quite small at older ages, we only focus on the data at age of $0-90$.
Therefore, there are $91$ columns each for Italy and Switzerland where each column represents an age group.
The mortality data are not available for Switzerland in $1872-1875$.
We use our proposed method to impute the missing mortality rates. 
\begin{figure}[!ht]
\caption{Spaghetti plots and heat maps for the mortality rate over age for Italy and Switzerland.
Black solid line represents the Spanish flu pandemic.
Dashed lines represent the World War I.
Dotted lines represent the World War II. 
Grey solid lines represent regular years.}
\label{FuncMort}
\begin{subfigure}{0.5\textwidth}
\centering
\caption{Mortality rate over age for Italy}
\label{MortAgeItaly}
\includegraphics[scale=0.25]{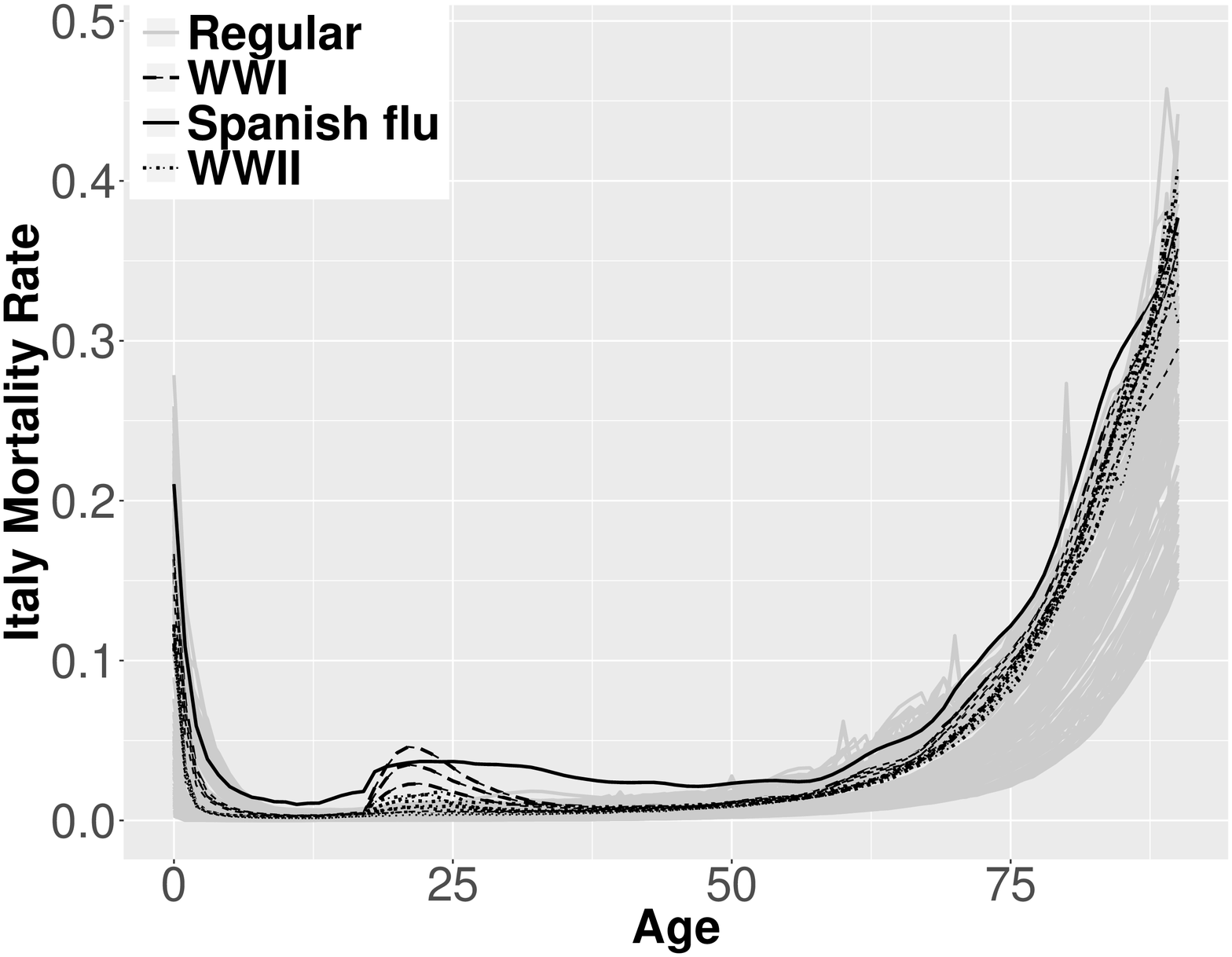}
\end{subfigure}
\begin{subfigure}{0.5\textwidth}
\centering
\caption{Mortality rate over age for Switzerland}
\label{MortAgeSwitz}
\includegraphics[scale=0.25]{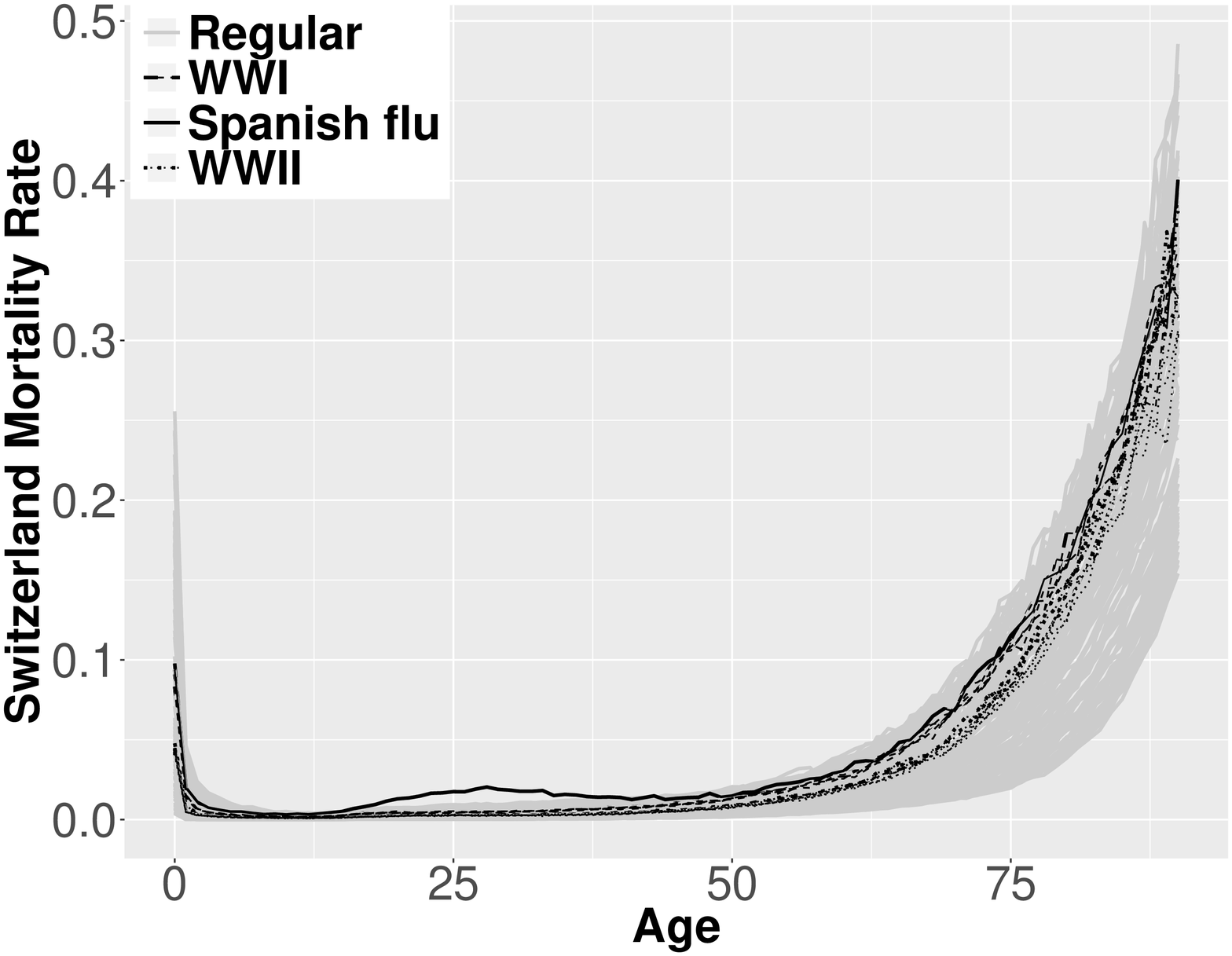}
\end{subfigure}\vfill
\begin{subfigure}{0.5\textwidth}
\centering
\caption{True Mortality Rate for Italy}
\label{TrueItaly}
\includegraphics[scale=0.3]{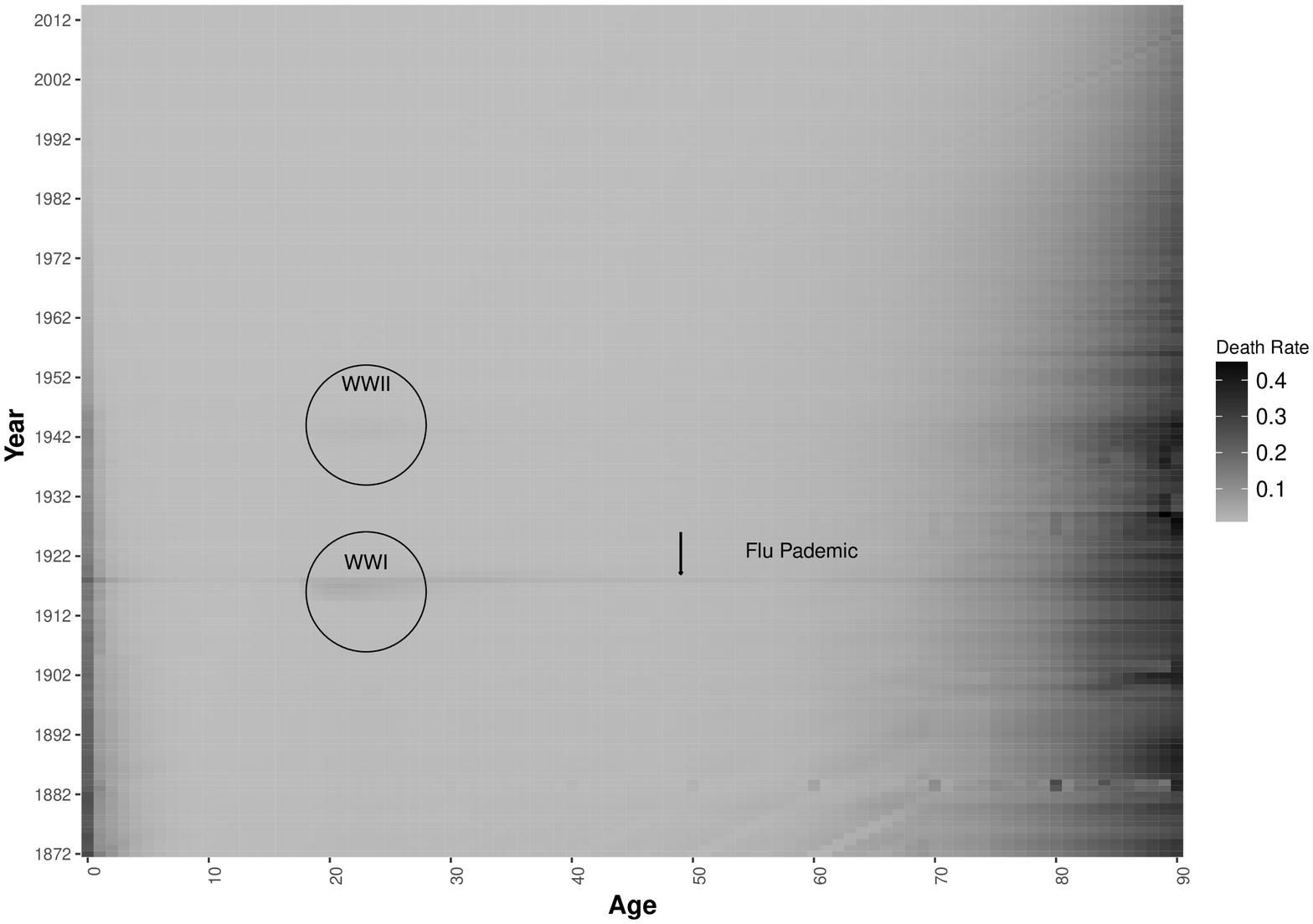}
\end{subfigure}
\begin{subfigure}{0.5\textwidth}
\centering
\caption{True Mortality Rate for Switzerland}
\label{TrueSwitz}
\includegraphics[scale=0.3]{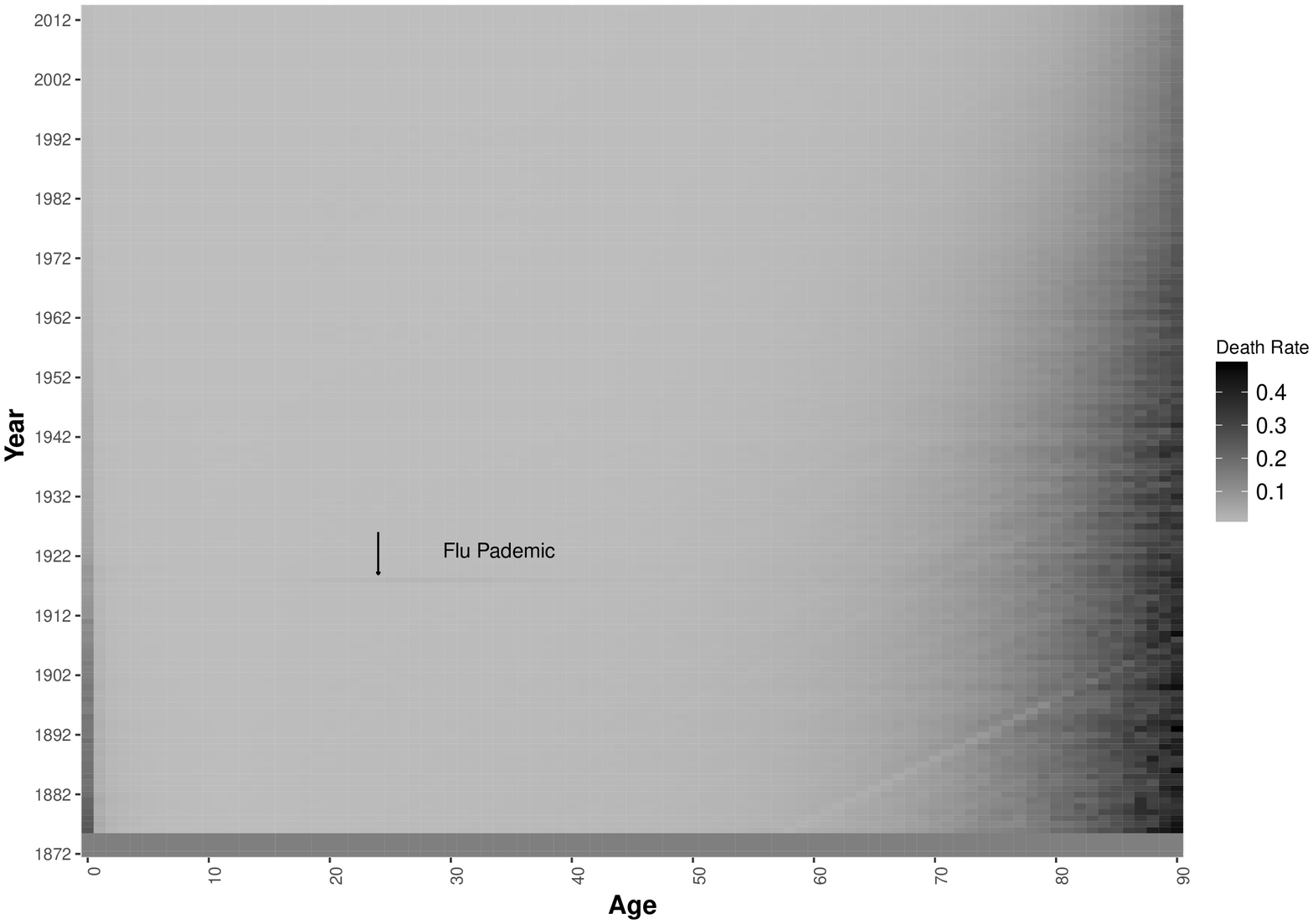}
\end{subfigure}
\end{figure}

Figure \ref{FuncMort} illustrates the mortality rates across age groups in different years in each country.
The mortality rates are calculated by taking the ratio of the number of exposure-to-risk and corresponding population size.
Figure \ref{MortAgeItaly} and Figure \ref{MortAgeSwitz} are the curve plots which show the mortality rate as a function of age and each curve represents a year.
They show that the mortality rate is relatively high at an early age and decreases dramatically after birth time.
The death rate remains stable after birth time to age $60$ and gradually increases after that time.
For Italy, several curves (dashed line and dotted line in Figure \ref{MortAgeItaly}) have a surge around 20 years old.
Those curves are mortality rate curves in the year from $1915$ to $1918$ and from $1942$ to $1944$, when World War I (WWI) and World War II (WWII) happened.
The two world wars led to a mass death of young adults in these years.
A curve (black solid) in Figure \ref{MortAgeItaly} and Figure \ref{MortAgeSwitz} stands out against the other curves across all age groups.
This black solid curve is the mortality rate curve in $1918$ when Spanish flu pandemic happened and led to a mass death for people of all age groups.
Figure \ref{TrueItaly} and Figure \ref{TrueSwitz} are the heat maps for the true mortality rate of Italy and Switzerland.
In the heat map for Italy, the two outlying periods are shown by two horizontal strips in Figure \ref{TrueItaly}.
The first strip around age $20$ is the period during the time of WWI.
Within this period, there is an outlying line across all the age groups, which is the time of Spanish flu pandemic.
The second strip around $20$ years old is the period during the time of WWII.
There is only one thin horizontal line in Figure \ref{TrueSwitz}, which is the period during the flu pandemic.

We apply GIPCA to the mortality in both countries.
First, we use BIC to estimate the ranks of the underlying structures.
By using the stepwise BIC algorithm, we reach to a rank estimation such that $r_J=13, r_1=7, r_2 =0$.
We check the trajectory of stepwise BIC values.
By comparing the BIC values of $r_J = 1, r_1 = 1, r_2 = 0$ and $r_J=13, r_1=7, r_2 =0$, we figure out that the improvement in BIC for the more complex decomposition is negligible.
Thus, we choose $r_J = 1, r_1 = 1, r_2 = 0$, which leads to a simple and intuitive decomposition of the raw data.

\begin{figure}[!ht]
\centering
\caption{Estimated result}
\label{fig:EstUV}
\begin{subfigure}{0.4\textwidth}
\centering
\caption{Estimated means for Italy}
\label{fig:MeanItaly}
\includegraphics[scale=0.2]{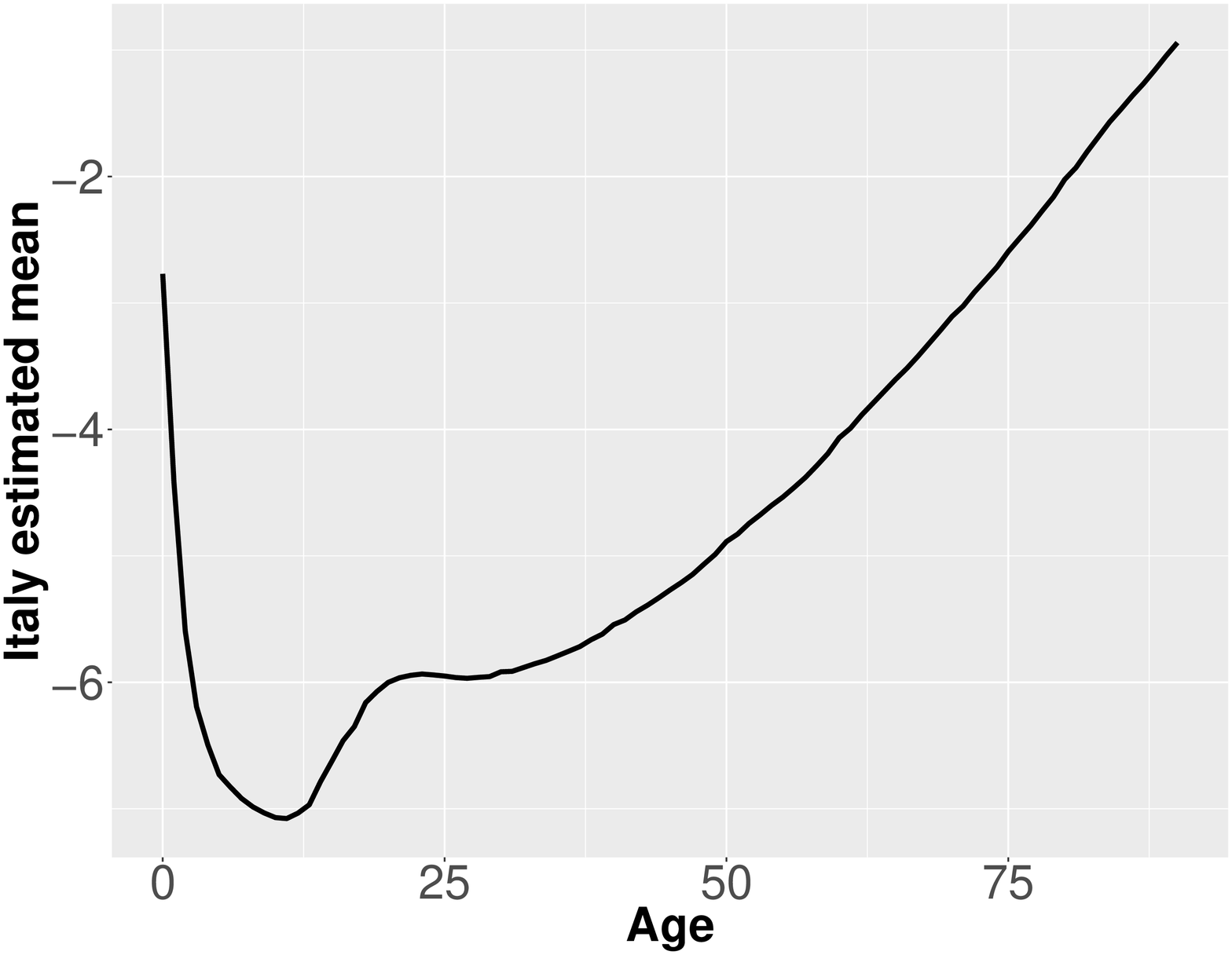}
\end{subfigure}
\begin{subfigure}{0.4\textwidth}
\centering
\caption{Estimated means for Switzerland}
\label{fig:MeanSwitz}
\includegraphics[scale=0.2]{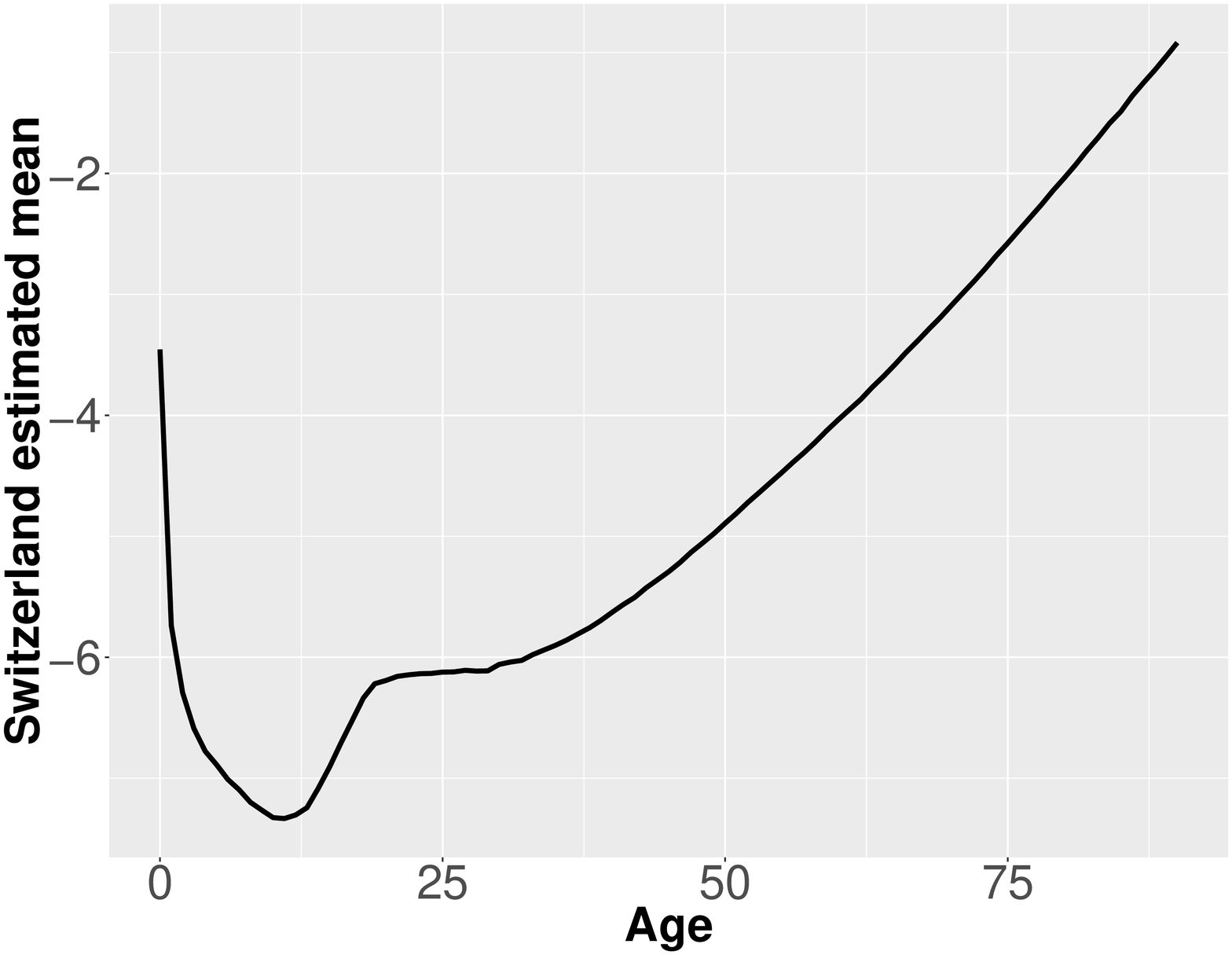}
\end{subfigure}

\begin{subfigure}{0.4\textwidth}
\centering
\caption{Estimated joint score matrix}
\label{fig:Ujoint}
\includegraphics[scale=0.2]{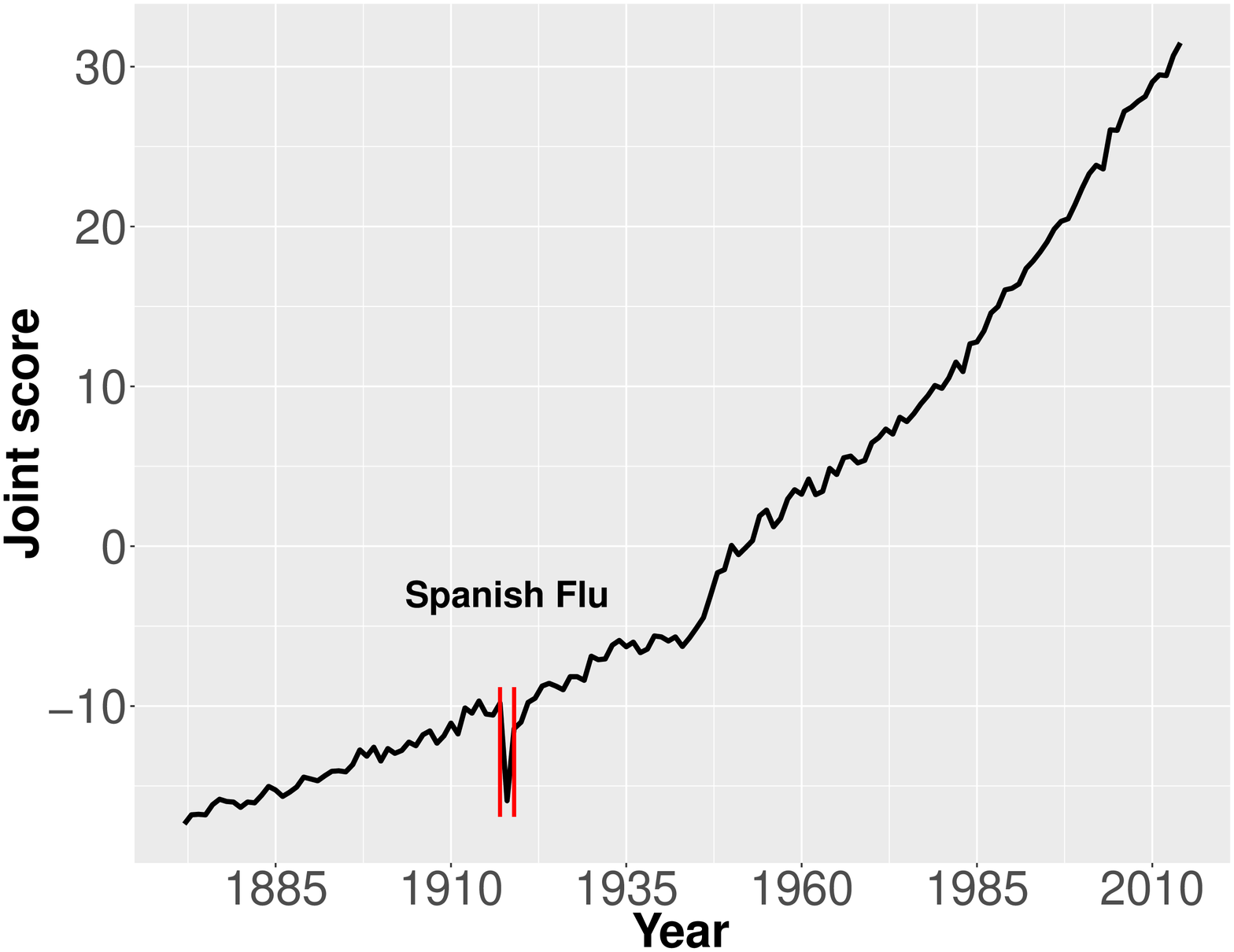}
\end{subfigure}\vfill
\begin{subfigure}{0.4\textwidth}
\centering
\caption{Estimated joint loading matrix for Italy}
\label{fig:VItaly}
\includegraphics[scale=0.2]{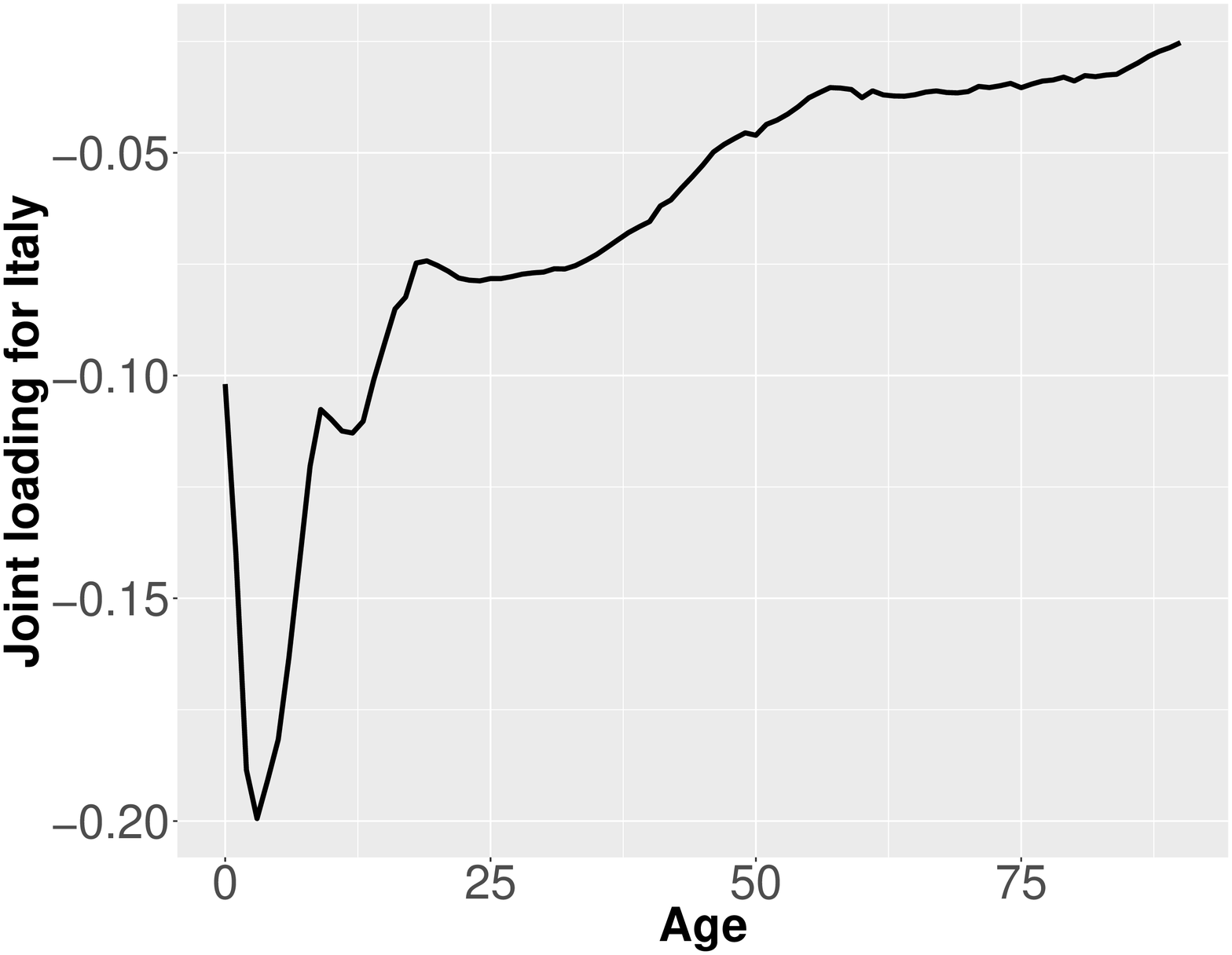}
\end{subfigure}
\begin{subfigure}{0.4\textwidth}
\centering
\caption{Estimated joint loading matrix for Switzerland}
\label{fig:VSwitz}
\includegraphics[scale=0.2]{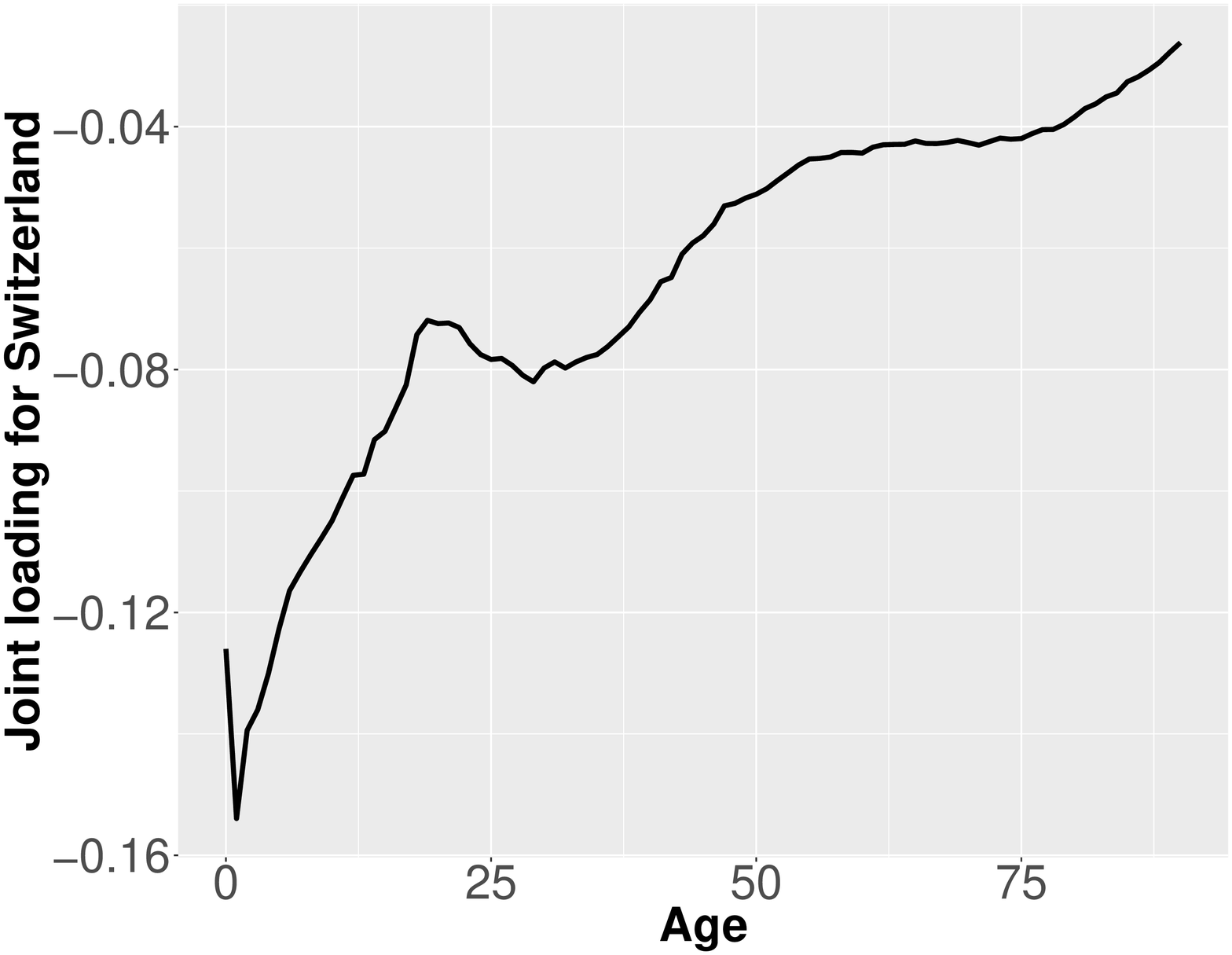}
\end{subfigure}
\begin{subfigure}{0.4\textwidth}
\centering
\caption{Estimated individual score matrix for Italy}
\label{fig:UindItaly}
\includegraphics[scale=0.2]{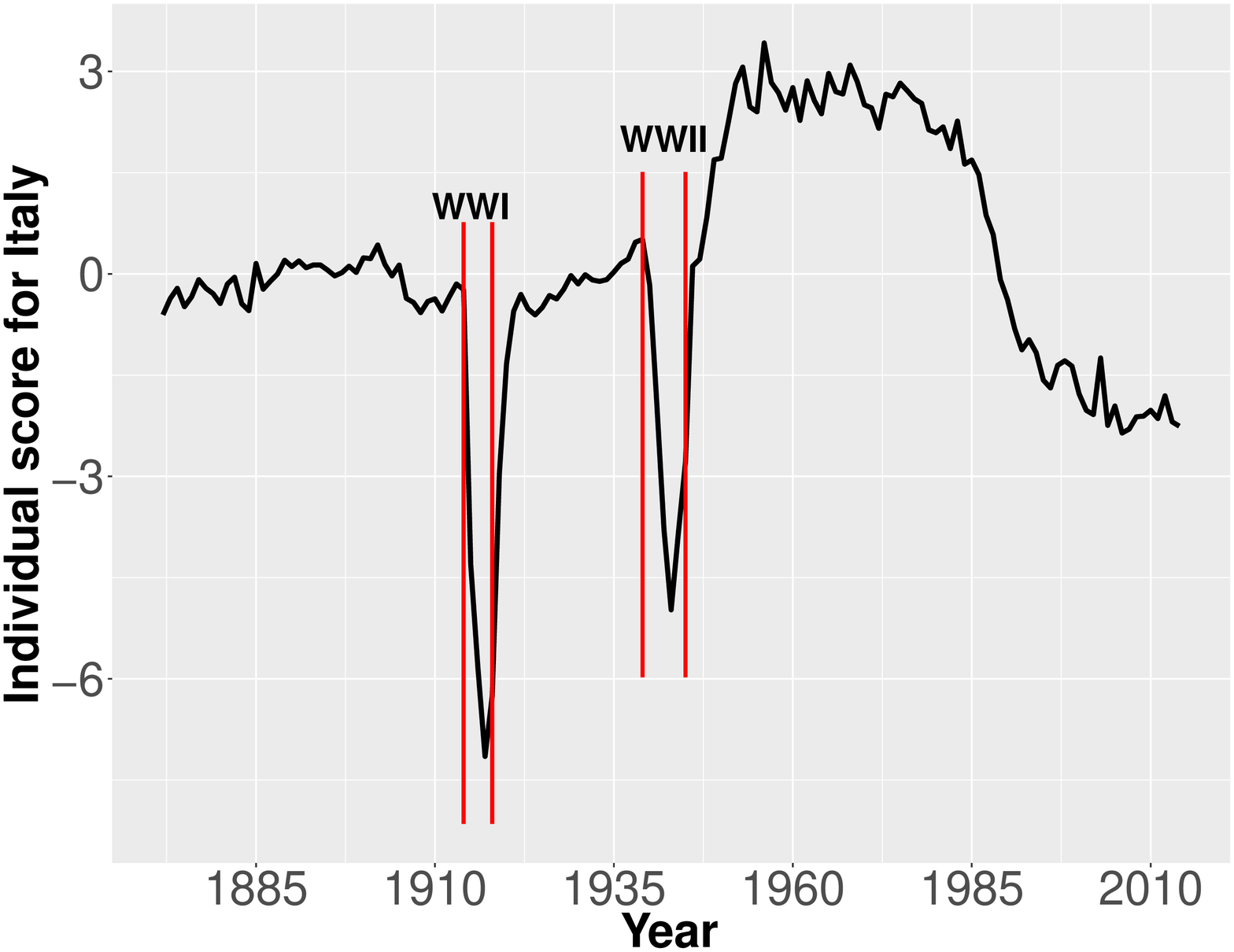}
\end{subfigure}
\begin{subfigure}{0.4\textwidth}
\centering
\caption{Estimated individual loading matrix for Italy}
\label{fig:AItaly}
\includegraphics[scale=0.2]{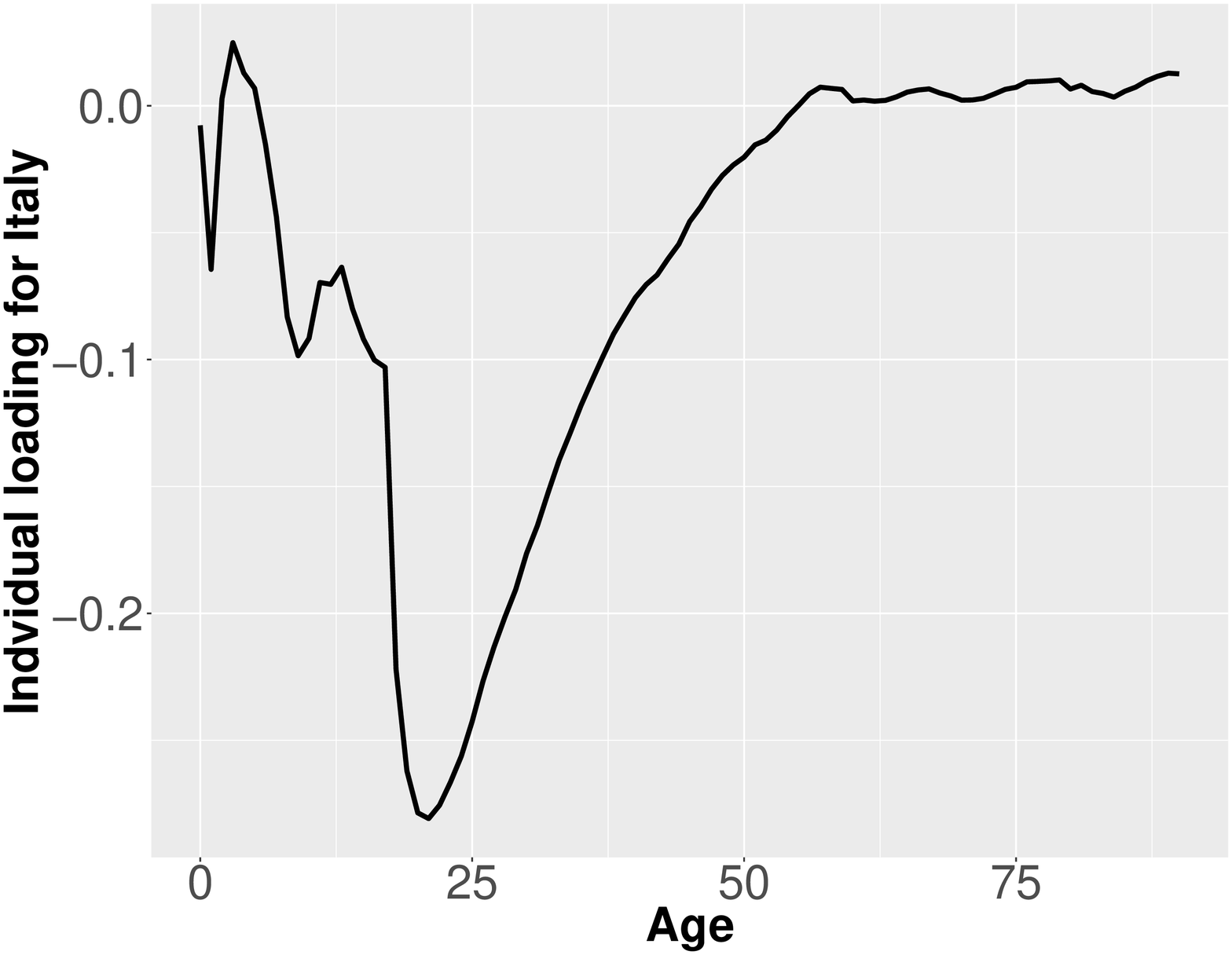}
\end{subfigure}
\end{figure}

The data types for both data sets are binomial.
The link functions for both data sets are logit function.
Following Algorithm \ref{alg:GIPCA} with rank $r_J = 1,r_1 = 1,r_2 = 0$, we get the estimates for the means, joint score and loading matrices, and individual score and loading matrices for Italy and Switzerland.
Figure \ref{fig:EstUV} visualizes the estimation results.
Figure \ref{fig:MeanItaly} and Figure \ref{fig:MeanSwitz} are the estimations of the column means for Italy and Switzerland correspondingly.
The two figures demonstrate the overall age-dependent component of mortality rate.
It decreases for early age groups and after a certain age, it increases exponentially.
The pattern agrees with the Gompertz--Makeham law of mortality, which states that the mortality rate consists of an age-independent component and age-dependent component, increasing with age exponentially \citep{GompertzMortality}.
Figure \ref{fig:Ujoint} illustrates the estimated left singular vector (score matrix) for joint structure (i.e., the shared time-varying pattern of mortality rates in different countries).
The score vector has a clear dip around year $1918$, which is the period of Spanish flu pandemic.

Figure \ref{fig:VItaly} and Figure \ref{fig:VSwitz} are the joint loading vector for the two countries.
The estimated loading vectors demonstrate that Spanish flu pandemic resulted in a mass death to younger people such as infants and teenagers.
The estimated individual score vector for Italy is shown in Figure \ref{fig:UindItaly}.
It has two apparent dips around $1917$ and $1943$, which correspond to the periods of World War I and World War II.
The individual loading vector for Italy is shown in Figure \ref{fig:AItaly}.
It shows that the population aged around $20$ to $25$ was mostly affected, probably because they were directly involved in the wars.
Switzerland remained neutral during both wars and therefore does not express this mortality pattern.

Next, we evaluate the imputation performance of the proposed method.
In particular, we consider three ad hoc methods for imputing missing mortality rates that are commonly used in practice: mean, adjacent years, and same year imputation.
More specifically,
\begin{itemize}[noitemsep,topsep=0.5pt]
\item[]\textbf{Ad Hoc 1 (Mean Imputation)}: The missing entries are imputed with the mean of mortality rate at the same age within the same data set.

\item[]\textbf{Ad Hoc 2 (Adjacent Years Imputation)}: The missing entries are imputed with the average of mortality rate within minus/plus $5$ years of the same age group within the same data set.

\item[]\textbf{Ad Hoc 3 (Same Year Imputation)}: The missing entries are imputed with the mortality rate in the same year of the other data set.

\end{itemize}
We randomly pick $10$ rows (i.e., years) for each data set and set them to be missing.
Our proposed method and $3$ ad hoc methods are applied to the block-wise missing data.
We repeat the procedure $100$ times.
The estimated mortality rates are the inverse logit of the estimated natural parameter matrix.
The block-wise missing entries are imputed by the inverse logit of the corresponding estimated joint structure.
We calculate $DiffR_{miss}$ (Model \eqref{LossMiss}) to the imputed and true data and running time for GIPCA and other approaches.

\begin{table}[!ht]
\centering
\caption{We randomly pick $10$ rows for each data set and set them to be missing.
Impute the block-wise missing for Italy and Switzerland using GIPCA and 3 ad hoc approaches.
The above procedure is repeated $100$ times.
The median and the median absolute deviation (MAD, in the parenthesis) for the relative Frobenius loss mentioned in section 4.2 are calculated.
The best results are highlighted in bold.}
\small
\label{MortImput}
\begin{tabular}{lll}
\hline\hline
Method & Italy & Switzerland\\
\hline
\textbf{GIPCA} & \textbf{0.137 (0.037)}&\textbf{0.084 (0.009)} \\
\textbf{Ad Hoc 1 Mean Imputation} & 0.314 (0.056) &0.319 (0.046)\\
\textbf{Ad Hoc 2 Adjacent Year Imputation} &0.468 (0.140)& 0.490 (0.127)\\
\textbf{Ad Hoc 3 Same Year Imputation} &0.163 (0.035) &0.164 (0.024)\\
\hline\hline
\end{tabular}
\end{table}

Imputation performance results in Table \ref{MortImput} show that GIPCA outperforms the other $3$ ad hoc approaches in terms of the relative Frobenius loss.
Among the $3$ ad hoc methods, the imputation accuracy for \textbf{Ad Hoc 3} is the closest to what we have for GIPCA.
The better performance validates the assumption we make for \textbf{Ad Hoc 3} that the mortality rates are similar between Italy and Switzerland in the same year.
On the other hand, it also agrees with the imputation mechanism implemented by GIPCA that we use the joint association to impute the missing entries.
\textbf{Ad Hoc 2} performs the worst among all the methods.
The unsatisfactory results of \textbf{Ad Hoc 1} and \textbf{Ad Hoc 2} indicate that simply using average across different years within one data set to impute the missing mortality is limited.
When we have two or more data sets, which share the same samples, imputing missing entries by taking the advantage of the shared traits among different data sets is better than using the average within one data set.

\section{Discussion}
\label{Sec:discuss}
In this paper, we develop a generalized integrative principal components analysis approach for dimension reduction of data sets from multiple sources with different data types.
Our proposed method is also able to deal with multi-source data sets containing block-wise missing entries.
We apply the proposed method to mortality data in Italy and Switzerland and identify some meaningful signals, and achieve good missing data imputation accuracy.
We also develop a rank selection approach derived from BIC, which accommodates multi-source data of different distributional types.

Based on the result in Section \ref{SimResult}, the stepwise BIC approach performs well in most scenarios.
However, when we have data from more than two sources,
the accuracy tends to lower.
Alternative rank selection methods call for more investigation.
As to the proposed algorithm, although the current  GIPCA algorithm only applies to the exponential family distributions, the general idea can be extended to more general non-Gaussian distributions.
Extensions to other distributions are future research directions.
\newpage
\bibliography{refs}
\bibliographystyle{unsrtnat}

\newpage
\begin{center}
\huge Supplementary Materials For "Generalized Integrative Principal Component Analysis for Multi-Type Data with Block-Wise Missing Structure"
\end{center}

\setcounter{equation}{0}
\setcounter{figure}{0}
\setcounter{table}{0}
\setcounter{page}{1}
\setcounter{section}{0}
\makeatletter
\renewcommand\thesection{\Alph{section}}
\renewcommand{\thefigure}{S\arabic{figure}}
\renewcommand{\thetable}{S\arabic{table}}
\renewcommand{\bibnumfmt}[1]{[S#1]}
\renewcommand{\citenumfont}[1]{S#1}

\section{Result of Rank Selection by Adapted BIC}
We apply the rank selection procedure mentioned in Section 3.2 in the manuscript to the data generated from corresponding distribution independently for $50$ times when the natural parameter matrix is fixed.
BIC criterion (Model (3.5) in the manuscript) is used to estimate ranks for each simulation scenarios with different missing rates.
We apply the proposed BIC criterion to all the scenarios with different missing rate $0\%,5\%$, $10\%$.
Overall, the adapted BIC criterion performs well for different settings (Table \ref{RankSelect}).
The stepwise selection procedure correctly identifies the true ranks for joint structure and individual structures almost all the times for scenarios with two data types with various missing rates.
We also apply the selection procedure to
scenario \textbf{Gaussian-Poisson-binomial}. 
The proposed adapted BIC criterion is unable to specify the rank combinations correctly (misspecified as $r_J = 3, r_{A_{Gaussian}} = r_{A_{Poisson}} = 1, r_{A_{binomial}} = 2$). 
This may be because the signal-to-noise ratio for the binomial data is relatively low compared to the other datasets.  Alternative approaches to rank selection that can accommodate to multiple ($>$2) sources of data call for more investigation.

\begin{table}[!ht]
\tablinesep=1ex
\centering
\caption{Rank selection result for Scenario 1 (Gaussian Gaussian), Scenario 2 (Gaussian Poisson), Scenario 3 (Gaussian binomial), and Scenario 5 (binomial Poisson) with different missing rates. The number of correctly specified ranks is out of $50$.}
\label{RankSelect}
\begin{tabular}{Sp{0.5cm}p{0.5cm}p{0.5cm}}
\hline\hline
& \multicolumn{3}{c}{\textbf{Missing Rate \%}}\\
\textbf{Scenario}& 0 & 5 &10\\
\hline
\textbf{Scenario 1}& 50 & 50 &50 \\
\textbf{Scenario 2} &49 & 49 &50 \\
\textbf{Scenario 3}&50 &50 &50 \\
\textbf{Scenario 5} & 50  & 50&49\\
\hline\hline
\end{tabular}
\end{table}
\section{Simulation Result for Gaussian Poisson binomial Scenario}
In addition to the scenarios with data sets from two sources with different data types, we also apply the propose approach to the scenario with three different data types, Gaussian, Poisson and binomial (\textbf{Scenario 4}).
The ranks for each part are set to be $2$.
Joint and individual score matrices $(\mathbf{U}_0,\mathbf{U}_1,\mathbf{U}_2,\mathbf{U}_3)$ are filled with uniform random numbers $Unif(-0.5,0.5)$ and normalized to have orthonormal columns.

\textbf{Scenario 4}: \textbf{Gaussian-Poisson-binomial}
The joint loading matrices $\boldsymbol{V}_1$ for Gaussian,$\boldsymbol{V}_2$ for Poisson are generated from $Unif(-0.5,0.5)$, and $\boldsymbol{V}_3$ for binomial is generated from $Unif(-1.5,1.5)$.
The individual loading matrices $\boldsymbol{A}_1$ (Gaussian), $\boldsymbol{A}_2$ (Poisson), $\boldsymbol{A}_3$ (binomial) are generated from $Unif(-0.5,0.5)$, $Unif(-0.25,0.25)$, and $Unif(-1.5,1.5)$ correspondingly.
The singular values of the joint structure are set to be $(300,280)$, the singular values of the individual structures to be $(150,120)$ for Gaussian, $(150,140)$ for Poisson and $(200,180)$ for binomial.
For such scenario, the imputation accuracy of GIPCA outperforms the other ad hoc methods for three data sets as well.

The means for Gaussian data set in each scenario that contains Gaussian data are generated from $Unif(-0.5,0.5)$.
The means of Poisson distribution to be positive (from $Unif(0,1)$) to mimic Poisson data in reality.
The means for binomial data set are generated from $Unif(-1.5,1.5)$.
For Gaussian data, we set the variance for the generated data to be $1$.
For binomial data, we set the number of trials to be $100$.
Similarly, as what is stated in Section 4.1,  data are generated from fixed natural parameter matrix.
The result in Table \ref{SimCompThree} shows that the proposed GIPCA outperforms the other ad hoc methods.
\begin{table}[!ht]
\centering
\caption{Simulation results for \textbf{Scenario 4} based on $100$ simulation runs when the natural parameter matrices are fixed for each data source.
The missing rate is 5\%.
The median and the median absolute deviation (MAD) for each evaluation criterion under each scenario are calculated.
MAD is in parenthesis.
The best results are highlighted in bold.}
\label{SimCompThree}
\begin{tabular}{ccccc}
\hline\hline
&\multicolumn{3}{c}{\textbf{Scenario 4}}\\
& \textbf{Gaussian} &\textbf{Poisson} &\textbf{binomial} & \textbf{Running Time}\\
\hline
\textbf{Adhoc1} &13.02 (8.06) &8.82 (4.03) & 7.10 (2.40) & 594.37 (148.9) \\
\textbf{Adhoc2} &4.60 (1.13)  &3.40 (1.31)  & 3.46 (0.98) & 3.38 (0.87)\\
\textbf{GIPCA} & \textbf{0.83 (0.00)}&\textbf{0.87 (0.00)} &\textbf{0.57 (0.00)} & 549.47 (119.39)\\
\hline\hline
\end{tabular}
\end{table}
\section{Simulation Results for  negligible joint structure}
If the joint structure is dominant, due to the reason that the imputation relies on the estimated joint structure, the missing imputation by our proposed approach would be more accurate. 
If the joint structure is negligible compared with the individual structure, our proposed approach can still handle such a situation.
However, since our proposed method directly exploits the joint structure for imputation, the imputation of the missing entries may not be accurate.
We explore the scenarios when the signals of the joint and individual structures are comparable in the manuscript (Table 1 in the manuscript and Table \ref{SimCompThree}).
In addition to the settings in the manuscript, we also explore the scenarios when the signals of the joint and individual structures are distinct.
In Table \ref{tbl:simresult}, we set the true singular values to construct the natural parameter matrix of the joint structure relatively small ($1/2$, $1/5$ or $1/10$ of the singular values in the original setting).
The results show that the performance of missing imputation for \textbf{Gaussian-Gaussian} and \textbf{Gaussian-Poisson} scenarios are relatively robust against the change of singular values.
For scenarios involving binomial distributions, the performance is sensitive to the change of signal.

\section{Sensitivity to Initial Values}
In order to evaluate how sensitive the algorithm to initial values, we set up different initial values in the proposed algorithm to fit the same two data sets.
The data set is generated the same as we described in Section 4.1.
For each scenario, we use the same simulated data, but we generate different initial values based on different random seed for the proposed algorithm.
Table \ref{tbl:sen} shows that the performance of missing imputation derived by the proposed method is stable, which indicates that our algorithm is not sensitive to initial values.
\begin{table}[!ht]
\centering
\caption{Simulation results for two data sets based on $100$ simulation runs when the natural parameter matrices are fixed for each data source.
A $r_J\times r_J$ matrix $\Sigma_J$ is a diagonal matrix whose diagonal elements are the singular values to construct the natural parameter matrix of the joint structure, where $r_J$ is the rank of joint structure.
The median and the median absolute deviation (MAD) for each evaluation criterion under each scenario are calculated.
MAD is in parenthesis.}
\label{tbl:simresult}
\small
\begin{tabular}{clllllll}
\hline
 &&\multicolumn{2}{c}{\textbf{Adhoc1}} &\multicolumn{2}{c}{\textbf{Adhoc2}}  & \multicolumn{2}{c}{\textbf{GIPCA}}\\
 && Source1 &Source2 &Source1 &Source2&Source1 &Source2\\
 \hline

\textbf{Gaussian Gaussian} & $DiffR_{Miss}$ &11.14 (0.85) & 8.41 (0.69) &  1.56 (0.00) & 1.00 (0.00) & 1.15 (0.06) & 0.93 (0.01)\\
$\Sigma_J/2$& \textbf{Running Time} & \multicolumn{2}{c}{100.45 (13.94)} & \multicolumn{2}{c}{3.26 (0.07)} & \multicolumn{2}{c}{ 209.37 (145.26)}\\

\textbf{Gaussian Gaussian} & $DiffR_{Miss}$ & 12.22 (1.06) & 9.13 (0.77) & 1.65 (0.00) & 1.00 (0.00) & 1.55 (0.01) & 1.00 (0.00)\\
$\Sigma_J/5$ & \textbf{Running Time} & \multicolumn{2}{c}{101.51 (14.45)} & \multicolumn{2}{c}{3.26 (0.04)} & \multicolumn{2}{c}{138.5 (26.69)}\\

\textbf{Gaussian Gaussian} & $DiffR_{Miss}$ &13.4 (1.37) & 8.76 (1.10) & 1.67 (0.00) & 1.00 (0.00) & 1.15 (0.23) & 4.51 (5.12)\\
$\Sigma_J/10$ & \textbf{Running Time}
 &\multicolumn{2}{c}{114.71 (18.98)} & \multicolumn{2}{c}{3.28 (0.06)} & \multicolumn{2}{c}{396.54 (295.22)} \\

\textbf{Gaussian Poisson} & $DiffR_{Miss}$ & 13.13 (4.98) & 6.01 (0.4) & 1.23 (0.05) & 1.64 (0.01) & 0.73 (0.00) & 0.74 (0.00)\\
$\Sigma_J/2$ & \textbf{Running Time}&\multicolumn{2}{c}{212.39 (35.78)} & \multicolumn{2}{c}{1.84 (0.05)}& \multicolumn{2}{c}{152.27 (35.57)}\\

\textbf{Gaussian Poisson} & $DiffR_{Miss}$ & 10.06 (6.39) & 7.03 (0.71) & 2.81 (0.43) & 1.25 (0.02) & 0.95 (0.01) & 0.99 (0.02) \\
$\Sigma_J/5$ & \textbf{Running Time} & \multicolumn{2}{c}{196.02 (37.53)} & \multicolumn{2}{c}{1.84 (0.06)} & \multicolumn{2}{c}{394.18 (138.01)}\\

\textbf{Gaussian Poisson} & $DiffR_{Miss}$ &  15.97 (6.19) & 8.02 (1.05) & 1.17 (0.00) & 1.64 (0.01) & 1.02 (0.04) & 6.91 (2.52)\\
$\Sigma_J/10$ & \textbf{Running Time} & \multicolumn{2}{c}{210.66 (42.03)} & \multicolumn{2}{c}{1.84 (0.10)} & \multicolumn{2}{c}{595.02 (196.2)}\\

\textbf{Gaussian binomial} & $DiffR_{Miss}$& 8.06 (3.40) &7.56 (1.72)& 1.04 (0.00) & 0.99 (0.00) & 0.95 (0.01) & 0.77 (0.05)\\
$\Sigma_J/2$ & \textbf{Running Time} & \multicolumn{2}{c}{502 (97.77)} & \multicolumn{2}{c}{1.78 (0.08)} & \multicolumn{2}{c}{835.26 (483.43)}\\

\textbf{Gaussian binomial} & $DiffR_{Miss}$ & 11.09 (5.37) & 1.43 (0.32) &  1.12 (0.00) & 2.05 (0) & 1.02 (0.01) &  6.54 (0.79)\\
$\Sigma_J/5$ & \textbf{Running Time} & \multicolumn{2}{c}{467.65 (244.06)} &\multicolumn{2}{c}{1.87 (0.13)} & \multicolumn{2}{c}{391.85 (87.33)}\\

\textbf{Poisson binomial} & $DiffR_{Miss}$ &  11.07 (5.16) & 13.26 (6.85) & 0.88 (0.06) & 5.83 (6.83) & 3.63 (2.44) & 0.99 (0.03)\\
$\Sigma_J/2$ &  \textbf{Running Time} & \multicolumn{2}{c}{190.22 (39.83)} & \multicolumn{2}{c}{0.39 (0.04)} & \multicolumn{2}{c}{ 272.48 (136.3)}\\

\hline
\end{tabular}

\end{table} 

\begin{table}[!ht]
\centering
\caption{Simulation results ($DiffR_{Miss}$) for the same two data sets generated by fixed natural parameter matrices for each source and repeated for $100$ time with different initial values for the algorithm.
The median and the median absolute deviation (MAD) for each evaluation criterion under each scenario are calculated.
MAD is in parenthesis.}
\label{tbl:sen}
\begin{tabular}{ccc}
\hline\hline
& \textbf{Simulated data I} & \textbf{Simulated data II}\\
\hline
\textbf{Gaussian Gaussain} & 0.65 (0.00) & 0.72 (0.00)\\
\textbf{Gaussian Poisson} &  0.46 (0.00)&0.47(0.00)\\
\textbf{Gaussian binomial} &0.77 (0.00)& 0.44(0.01)\\
\textbf{Poisson binomial} & 0.59 (0.01) & 0.84 (0.00)  \\
\hline\hline
\end{tabular}
\end{table}

\newpage

\clearpage

\end{document}